\documentclass[a4paper,fleqn,usenatbib,useAMS]{mnras}

\usepackage[T1]{fontenc}
\usepackage{ae,aecompl}

\usepackage{graphicx}	
\usepackage{amsmath}	
\usepackage{amssymb}	
\usepackage{float}

\newcommand\dd{\mathrm{d}}

\newcommand\ee{\mathrm{e}}
\newcommand\ii{\mathrm{i}}

\title[On the flow induced by a low-mass planet]
{Horseshoes and spiral waves: capturing the 3D flow induced by a low-mass planet analytically}

\author[Joshua J. Brown and Gordon I. Ogilvie]
{Joshua J. Brown\thanks{E-mail: jb2228@cam.ac.uk} and Gordon I. Ogilvie\thanks{E-mail: gio10@cam.ac.uk}
\\
Department of Applied Mathematics and Theoretical Physics,
University of Cambridge, Centre for Mathematical Sciences,\\
Wilberforce Road, Cambridge CB3 0WA, UK
}

\date{Accepted 2024 August 30. Received 2024 August 30; in original form 2024 June 19}

\pubyear{2024}

\begin{document}
\label{firstpage}
\pagerange{\pageref{firstpage}--\pageref{lastpage}}
\maketitle

\begin{abstract}
The key difficulty faced by 2D models for planet-disc interaction is in appropriately accounting for the impact of the disc's vertical structure on the dynamics. 3D effects are often mimicked via softening of the planet's potential; however, the planet-induced flow and torques often depend strongly on the choice of softening length. We show that for a linear adiabatic flow perturbing a vertically isothermal disc, there is a particular vertical average of the 3D equations of motion which exactly reproduces 2D fluid equations for arbitrary adiabatic index. There is a strong connection here with the Lubow-Pringle 2D mode of the disc. Correspondingly, we find a simple, general prescription for the consistent treatment of planetary potentials embedded within ‘2D’ discs. The flow induced by a low-mass planet involves large-scale excited spiral density waves which transport angular momentum radially away from the planet, and ‘horseshoe streamlines’ within the co-orbital region. We derive simple linear equations governing the flow which locally capture both effects faithfully simultaneously. We present an accurate co-orbital flow solution allowing for inexpensive future study of corotation torques, and predict the vertical structure of the co-orbital flow and horseshoe region width for different values of adiabatic index, as well as the vertical dependence of the initial shock location. We find strong agreement with the flow computed in 3D numerical simulations, and with 3D one-sided Lindblad torque estimates, which are a factor of 2 to 3 times lower than values from previous 2D simulations.
\end{abstract}




\section{Introduction}
\label{s:intro}

Discs of gas in orbital motion around a massive central body are found in many astronomical contexts. Objects embedded within these discs, for example planets embedded in circumstellar discs, or stellar-mass black holes within AGN discs, are in general subject to strong gravitational interactions with the gas and dust which comprise the discs. Indeed, signatures such as gaps, rings and spirals observed in discs' emission are highly suggestive of large embedded planets \citep{benisty_asymmetric_2015,grady_spiral_2013,alma_partnership_2014_2015,andrews_disk_2018,andrews_observations_2020}. These interactions lead to angular momentum exchange between the object and the disc, causing the object to migrate.

Type I migration concerns low-mass embedded objects which excite predominantly linear perturbations to the flow in the disc. Not only is it a crucial ingredient in theory of planet formation and population synthesis \citep{mordasini_planetary_2018}, but it also may play a key role in understanding the rapid merger rate of stellar-mass black hole (sBH) binaries inferred from LIGO detections \citep{secunda_orbital_2019}. Whilst the migration of sBHs within very thin AGN discs is arguably better suited to the type I parameter regime, the development of type I migration theory has been driven by the need to understand and predict the dynamics of young planets.

Linear two-dimensional analyses from early work \citep{goldreich_excitation_1979,goldreich_disk-satellite_1980,artymowicz_wave_1993,korycansky_numerical_1993} presented a picture in which the planet exchanged angular momentum with the disc via two mechanisms: some angular momentum was transported away from the planet by density waves excited at Lindblad resonances, and some separately `absorbed' into corotation resonances. The torque depends only on the properties of the disc and their gradients near the planet, since it is exerted predominantly locally. Type I torque formulae therefore take very simple forms. \citet{tanaka_threedimensional_2002}, and more recently \citet{tanaka_three-dimensional_2024}, extended the 2D linear calculation to 3D locally isothermal discs, arriving at a torque formula in good agreement with the numerical simulations of \citet{dangelo_three-dimensional_2010}, which drives rapid inward migration on a time-scale of $\sim 10^5$ years for an Earth-mass planet.

This unopposed rapid inward migration is worrying; it acts as a potential obstacle to the feasibility of the core accretion model, in which giant gas planets form via the accretion of gas onto a solid core on a time-scale of $\gtrsim 10^6$ years \citep{lissauer_models_2009}. Furthermore, since the lifetime of the disc is $\lesssim 10^7$ years, it suggests at surface level an additional paradoxical existential threat for extra-solar Earth-like planets. Indeed, planetary population synthesis models have struggled to reproduce the observed distribution of semi-major axes among detected low-mass exoplanets from the linear type I torque estimates \citep{mordasini_extrasolar_2009}.

However, as pointed out by \citet{paardekooper_corotation_2009,paardekooper_width_2009}, the linear treatment of the corotation resonance is inappropriate, unless the disc is sufficiently viscous. Close to corotation, the radial displacement of fluid elements becomes non-linear, as they execute ‘horseshoe turns’ (akin to the behaviour of test particles seen in \citet{dermott_dynamics_1981} and \citet{petit_satellite_1986}), exchanging angular momentum with the planet as they are excited onto inner or outer orbits. An asymmetry between the potential vorticity (PV), or vortensity, on incident inner and outer orbits leads to a net torque comparable to that of the Lindblad torque \citep{ward_horsehoe_1991}, not captured in the linear theory.

In addition, the longer time-scale associated with this horseshoe motion means that additional physical effects become relevant, each with an associated strong torque with potential to resolve the aforementioned `paradoxes'. This has motivated a wealth of research, both semi-analytical and computational, to quantify the dependence of the corotation torque on diffusive, radiative and migration-feedback processes, among others \citep{balmforth_non-linear_2001,masset_coorbital_2001,masset_runaway_2003,paardekooper_halting_2006,paardekooper_disc_2008,paardekooper_corotation_2009,paardekooper_width_2009,masset_horseshoe_2009,masset_saturated_2010,paardekooper_torque_2010,paardekooper_torque_2011,lega_migration_2014,masset_horseshoe_2016,masset_coorbital_2017,fung_3d_2015,mcnally_low-mass_2020,paardekooper_planet-disk_2022}.

The torques which arise often depend strongly on the precise geometry of the coorbital flow, and correspondingly the value of the softening length, $b$ (often used to modify the planetary potential to mimic 3D effects in 2D disc models). The variation of torque components can be by more than a factor of 3 for different reasonable choices of $b$ (e.g. \citet{tanaka_three-dimensional_2024}, figure 5); such varied behaviour for different softening lengths is an issue not limited to the study of planetary torques \citep{muller_treating_2012}. In this paper we aim to accurately describe the flow induced by the planet, on top of which torque-inducing physics may be studied. Though there is much to be achieved and understood here, the coorbital flow has received little analytical attention since \citet{paardekooper_width_2009}, despite their analysis suffering some limitations (which we point out in section \ref{s:reduction}).

A key objective of this work is to address the problem of capturing 3D disc physics with 2D equations. In particular, for low-mass objects embedded within vertically isothermal discs which permit adiabatic perturbations, we show that there exists a particular vertical average of the equations of motion which reproduces the commonly adopted 2D fluid equations. This averaging procedure is closely related to the operator which projects onto the Lubow-Pringle 2D mode \citep{lubow_wave_1993}. This paper is concerned with the non-axisymmetric generalisation of this mode. The 2D mode is so-called as it has the property $v_z = 0$, but in general it is $z$-dependent. It is a member of the wider family of 3D disc modes and describes the spiral wake as well as the horseshoe motion within the coorbital region. Most importantly, our averaging process yields a simple, general prescription for the consistent treatment of planetary potentials embedded within `2D' discs, as well as an interpretation for 2D models.

A quantitative description and understanding of the 2D mode of the planet-induced flow, which captures the impact of the disc's vertical extent on the spiral wake, also has important observational applications. Detection of disc-embedded exoplanets via their kinematic signatures is a promising method for finding very young protoplanets during their formation stages \citep{pinte_kinematic_2023}. As \citet{fasano_planet-driven_2024} point out, 3D effects impact the precise structure of the planet's spiral wake, with important consequences for the observational analysis of this signature, and consequently planet mass estimates.

In section \ref{s:governingeqs}, we project the 3D equations of motion onto this 2D flow. We derive (without further approximation) independent, second order, linear equations governing the 2D flow in section \ref{s:reduction}. In section \ref{s:results}, we present the flow solution. We discuss our findings in section \ref{s:discussion}, and draw our conclusions in section \ref{s:conclusions}.

\section{Governing equations}
\label{s:governingeqs}

For the remainder of the paper, we will assume two quantities to be small. Firstly, we assume the disc's aspect ratio $h = H/r$ to be much smaller than 1. Secondly, we assume the planet's mass $M_p$ to be much smaller than the `thermal mass' \citep{goodman_planetary_2001}. That is,
\begin{equation}
    q \equiv \frac{M_p}{M_\star} \ll \frac{M_\text{th}}{M_\star} = h^3,
\end{equation}
where $M_\star$ is the mass of the central star. This ensures that the equations governing the flow are almost everywhere well approximated as linear (this may be verified post hoc), and that the disc's vertical structure is almost everywhere determined by the star's gravity. We therefore have 2 small parameters, namely $h$ and $\frac{q}{h^3}$; we'll exploit the fact that both are small to make analytical progress.

\subsection{Unperturbed state}

Before continuing, it's important to define precisely the model for the disc with which our planet will interact. We will consider the planet to introduce a perturbation to the background disc structure outlined below.

Our background disc is steady, axisymmetric and comprised entirely of an ideal gas. It is in orbit about a star of mass $M_\star$, fixed at the centre of our frame, and is inviscid and non-self-gravitating to a first approximation. The gas which comprises it solves the steady Euler equations and ideal gas equation,
\begin{subequations}\label{Eulereq}
\begin{align}
&{\bf{u}}_0\cdot \nabla {\bf{u}}_0 = - \frac{1}{\rho_0} \nabla p_0 - \nabla \Phi_0,\label{eqn1a}\\
&\nabla \cdot \left(\rho_0 {\bf{u}}_0\right) = 0,\label{eqn1b}\\
&\frac{p_0}{\rho_0} = \frac{k_B}{\bar{\mu}}T_0,\label{eqn1c}
\end{align}
\end{subequations}
where ${\bf{u}}_0$, $\rho_0$, $p_0$ and $T_0$ are the velocity, density, pressure and temperature of the background disc. In addition, $k_B$ is Boltzmann's constant, and $\bar{\mu}$ the mean molecular mass. We introduce the cylindrical coordinates $(r, \theta, z)$, with the $z$-axis normal to the plane of the disc, so that 
\begin{equation}
\Phi_0 = -\frac{G M_\star}{\sqrt{r^2+z^2}}
\end{equation}
is the gravitational potential of the central star, and the velocity of the background state may be written as ${\bf{u}}_0 = r\Omega(r,z) \bf{e}_\theta$.

We assume the background disc's temperature, $T_0$, to be a prescribed function of $r$ only, the result of a relatively fast thermal relaxation in the vertical direction compared to the long time-scale of the evolution of the disc, and ignoring the hotter irradiated outer layers. We remark that the background disc need only be `locally isothermal' to a first approximation for the analysis performed in this paper to hold. Though this background state is idealised, the calculation we perform is robust: we define perturbations relative to the exact background state (though we are perhaps ignorant of its precise structure), but only need knowledge of its leading order behaviour in order to evaluate these perturbations.

Furthermore, the isothermal prescription applies only to the background state: we'll consider the planet to perturb this background state \emph{adiabatically}. This represents a good approximation provided the time-scale for thermal relaxation is longer than the planet's orbital period. We discuss the validity of this assumption in more detail in section \ref{s:disctherm}. We define the isothermal sound speed, $c_s(r)$, via
\begin{equation}
    c_s^2 = \frac{k_B}{\bar{\mu}}T_0.
\end{equation}
We define the scale height, $H$, and aspect ratio, $h$ via
\begin{equation}
    h \equiv \frac{H}{r} \equiv \frac{c_s}{r\Omega_K},
\end{equation}
where $\Omega_K$ is the Keplerian angular frequency, satisfying
\begin{equation}
 \frac{G M_\star}{r^3} = \Omega_K^2.
\end{equation}
We assume $h \ll 1$, so that the disc is thin. The angular velocity of the gas in the unperturbed disc satisfies
\begin{equation}
    \Omega = \Omega_K\left(1 + \mathcal{O}(h^2)\right).
\end{equation}
The density and pressure then satisfy
\begin{equation}
    \rho_0 = \frac{p_0}{c_s^2} = \frac{\Sigma_0(r)}{\sqrt{2\upi} H}\exp{\left(-\frac{z^2}{2H^2}\right)},
\end{equation}
where $\Sigma_0(r)$ is the surface density.

\subsection{Perturbation equations}

We now introduce a planet of mass $M_p$ on a Keplerian circular orbit of radius $r_p$ and angular frequency $\Omega_p = \Omega_K(r_p)\sqrt{1+q}$ to the background disc. We consider the perturbation problem in the frame corotating with the planet, and assume further that the flow in this frame is steady. The fluid velocity in this frame is ${\bf{v}} = {\bf{u}} - r\Omega_p {\bf{e}}_\theta$, and the relevant Euler equations become
\begin{subequations}\label{fullEulereq}
\begin{align}
&{\bf{v}}\cdot \nabla {\bf{v}}  + 2{\bf{\Omega}}_p \times {\bf{v}} = - \frac{1}{\rho} \nabla p - \nabla \Phi_t - \nabla \Psi_p,\label{eqnEFa}\\
&\nabla \cdot \left(\rho {\bf{v}}\right) = 0,\label{eqnEFb}\\
&{\bf{v}} \cdot \nabla \left(p \rho^{-\gamma}\right) = 0,\label{eqnEFc}
\end{align}
\end{subequations}
where $\Phi_t = \Phi_0(r,z) - \frac{1}{2}\left(r^2- 3r_p^2\right)\Omega_p^2$ is the tidal potential, and the planet's potential, including the indirect term arising from the acceleration of the frame centred on the star, is given by
\begin{equation}
    \Psi_p = - \frac{G M_p}{\sqrt{r_p^2 + r^2 - 2 r r_p \cos{\theta} + z^2}} + q r_p \Omega_p^2  r \cos{\theta}.
\end{equation}
We now define local quasi-Cartesian coordinates $x = r - r_p$, $y = r_p\theta$, and perturbed variables
\begin{subequations}
\begin{align}
    &\rho' = \rho - \rho_0,\\ &p' = p - p_0,\\ &v'_x = v_r,\\ &v'_y = v_\theta - r(\Omega - \Omega_p) = u_\theta - r \Omega,\\ &v'_z = v_z.
\end{align}
\end{subequations}
It's useful to define further the time-derivative following the orbital shear flow
\begin{equation}
D = -\frac{3}{2}x\Omega_p\upartial_y,
\end{equation}
and upon subtraction of the background state solution from the Euler equations (\ref{fullEulereq}), we find the balance at the next order in $h$ (assuming $x = \mathcal{O}\left(H\right)$ and $y = \mathcal{O}\left(H\right)$) is given by
\begin{subequations}\label{pertEulereq}
\begin{align}
&D v'_x - 2\Omega_p v'_y + \frac{\upartial_x p'}{\rho_p} = -\upartial_x \phi_p,\label{eqnEPa}\\
&D v'_y + \frac{1}{2}\Omega_p v'_x + \frac{\upartial_y p'}{\rho_p} = -\upartial_y \phi_p,\label{eqnEPb}\\
&D v'_z + \Omega_p^2 z \frac{\rho'}{\rho_p} + \frac{\upartial_z p'}{\rho_p} = -\upartial_z \phi_p,\label{eqnEPc}\\
&D \rho' + \rho_p\upartial_xv'_x + \rho_p\upartial_yv'_y + \upartial_z\left(\rho_pv'_z\right) = 0,\label{eqnEPd}\\
&D\left(p' - \gamma c_p^2\rho'\right) + (\gamma - 1)\Omega_p^2 z \rho_p v'_z = 0,\label{eqnEPe}
\end{align}
\end{subequations}
where $c_p = c_s(r_p)$,
\begin{equation}\label{philoc}
    \phi_p = -\frac{G M_p}{\sqrt{x^2+y^2+z^2}},
\end{equation}
and we have taken the leading order approximation to the background disc's density,
\begin{equation}
    \rho_{p}(z) \equiv \frac{\Sigma_p}{\sqrt{2\upi} H_p}\exp{\left(-\frac{z^2}{2H_p^2}\right)},
\end{equation}
for $\Sigma_p = \Sigma_0(r_p)$ and $H_p = H(r_p)$. The system of equations (\ref{pertEulereq}) describes locally the flow excited by a planet embedded in a stratified 3D disc. We could in theory relax the assumption $y \lesssim H$ by reintroducing the azimuthally global expression for the planet's potential, $\Psi_p$, in place of $\phi_p$, and applying periodic boundary conditions at $y = \pm \upi r_p$. The corresponding correction however is only of relative size $\mathcal{O}\left(h\right)$, so that local system involving $\phi_p$ (which we adopt for the remainder of the paper) becomes exact in the limit $h \to 0$.

The system (\ref{pertEulereq}) governs the excitation of the spiral density waves by the planet, permits downstream gravity waves (e.g. \citet{lubow_analytic_2014}), and also describes the horseshoe trajectories followed by fluid elements close to the planet's orbital radius. As noted by several authors, for example by \citet{tanaka_threedimensional_2002}, the density wave excitation is confined to the region extending only a few scale heights from the planet, though (\ref{pertEulereq}) does not capture the asymmetry between the inner and outer spiral wakes. Importantly, as we'll demonstrate in the next section, (\ref{pertEulereq}) also has the elegant property that it may be vertically integrated to derive exact 2D flow equations.

\subsection{Projection onto a 2D flow}
\label{s:proj}

Remarkably, under a particular choice of vertical averaging, the system of equations (\ref{pertEulereq}) may be transformed exactly into the familiar linearized 2D flow equations. Associated with this averaging procedure is a definite choice for the `softening' of the planetary potential (specified in equation (\ref{Phidefn})).

The averaging procedure defined below is closely related to the operator which projects onto the 2D mode of the disc found by \citet{lubow_wave_1993}. Indeed, the 2D equations we find govern the radial and azimuthal evolution of the amplitude of this mode. The 2D mode is a member of a larger family of 3D modes, it exists for $\gamma < 2$, and is often referred to as `2D', since it has the property that $v'_z = 0$, despite possessing a vertical dependence.

Before we proceed, it's helpful to first introduce the adiabatic sound speed, scale height and aspect ratio as
\begin{equation}
    c_\gamma = \sqrt{\gamma} c_p, \quad H_\gamma = c_\gamma/\Omega_p, \quad h_\gamma = H_\gamma/r_p.
\end{equation}
We combine (\ref{eqnEPd}) and (\ref{eqnEPe}) to eliminate $\rho'$:
\begin{multline*}
    c_\gamma^2 \times (\text{\ref{eqnEPd}}) + (\text{\ref{eqnEPe}})\\
    \implies D p' + c_\gamma^2 \rho_p\left(\upartial_x v'_x + \upartial_y v'_y\right) \\ 
    + (\gamma - 1)\Omega_p^2 z \rho_p v'_z + c_\gamma^2\upartial_z\left(\rho_pv'_z\right) = 0
\end{multline*}
\begin{multline}\label{vertdec1}
\implies D \frac{p'}{\rho_p} + c_\gamma^2 \left(\upartial_x v'_x + \upartial_y v'_y\right) +\\ c_\gamma^2 \exp{\left(\tfrac{z^2}{2 H_\gamma^2}\right)}\upartial_z\left[ v'_z \exp{\left(-\tfrac{z^2}{2 H_\gamma^2}\right)}\right] = 0.
\end{multline}
We now multiply (\ref{vertdec1}) by the factor $\exp{\left(-\tfrac{z^2}{2 H_\gamma^2}\right)} \propto \rho_p^{1/\gamma}$ and integrate. (Incidentally, the function $\rho_p^{1/\gamma}$ is the product of the background density distribution and the vertical profile of the 2D mode. In this way, the vertical integration may formally be seen to be an inner product with the 2D mode.) The result of the integration is the 2D mass conservation analogue
\begin{equation}\label{2DMC}
D P' + \Sigma_p c_\gamma^2\left(\upartial_x \bar{v}_x + \upartial_y \bar{v}_y\right) = 0,
\end{equation}
where we have defined
\begin{equation}\label{eqnavg}
    \bar{v}_x \equiv \left\langle v'_x\right\rangle, \quad \bar{v}_y \equiv \left\langle v'_y\right\rangle, \quad P' \equiv \Sigma_p \left\langle p'/\rho_p(z)\right\rangle,
\end{equation}
for vertical average $\left\langle\,\cdots\right\rangle$ defined via:
\begin{equation}\label{avg}
    \left\langle X \right\rangle \equiv \frac{1}{\sqrt{2\upi}H_\gamma}\int_{-\infty}^{\infty}X\exp{\left(-\frac{z^2}{2H_\gamma^2}\right)}\dd z.
\end{equation}
We may similarly apply the averaging procedure $\left\langle\,\cdots\right\rangle$ defined in (\ref{avg}) to equations (\ref{eqnEPa}) and (\ref{eqnEPb}) to obtain the 2D momentum equations
\begin{subequations}
\begin{align}
&D\bar{v}_x - 2\Omega_p \bar{v}_y + \frac{1}{\Sigma_p}\upartial_x P' = -\upartial_x \Phi_p,\label{2Dxmom}\\
&D\bar{v}_y + \frac{1}{2}\Omega_p \bar{v}_x + \frac{1}{\Sigma_p}\upartial_y P' = -\upartial_y \Phi_p,\label{2Dymom}
\end{align}
\end{subequations}
where now $\Phi_p$ is precisely defined as $\left\langle \phi_p \right \rangle$, that is,
\begin{equation}\label{Phidefn}
    \Phi_p \equiv \left\langle -\frac{G M_p}{\sqrt{x^2 + y^2 + z^2}} \right\rangle = -\frac{G M_p}{H_\gamma}\frac{\ee^{\frac{1}{4}s^2}}{\sqrt{2\upi}}K_0\left(\tfrac{1}{4}s^2\right),
\end{equation}
with $s = \sqrt{x^2 + y^2}/H_\gamma$, and $K_0(z)$ the modified Bessel function. Note that in a global 2D disc model, the above expression remains valid upon redefining $s = \left|{\bf{r}} - {\bf{r}}_p\right|/H_\gamma$ and reintroducing the indirect term (this matter is discussed in greater detail in section \ref{s:softening}). This prescription for the potential is therefore generally applicable to 2D disc models. Reassuringly, at large distances, the above expression becomes
\begin{equation}\label{phifar}
    \Phi_p = -\frac{G M_p}{\sqrt{x^2 + y^2}}\left(1 + \mathcal{O}\left(\frac{H_\gamma^2}{x^2+y^2}\right)\right),
\end{equation}
and close to the planet, our 2D potential has a logarithmic singularity. We therefore have exact 2D equations forced by a 2D potential, valid for arbitrary adiabatic index $\gamma$.

We remark that we did not make use of equation (\ref{eqnEPc}) for the vertical velocity; the 2D mode is orthogonal to and ignorant of the permitted vertical motions of the disc (including downstream gravity waves and inertial waves). Studies of these excited gravity waves indicate they may have an important observational signature and impact on the torque on the planet, which is difficult to predict due to the intricacy of the problem \citep{zhu_planet-disk_2012,lubow_analytic_2014,mcnally_low-mass_2020,bae_observational_2021,ziampras_buoyancy_2023}. That being said, much of the important dynamics are captured by the 2D mode, and we shall focus the majority of our attention here for the remainder of the paper.

\subsection{Relation to 2D Euler equations}\label{s:2D}

2D disc models enjoy a valuable simplicity compared to 3D models. However, the non-linear 2D Navier-Stokes equations comprise only an approximate model for the flow in the disc. The analysis performed above (as well as that in section \ref{s:td}) provides an interpretation for the 2D flow variables in terms of their counterparts in a 3D disc with vertical extent. This relationship holds when the disc's response (for example to forcing by a perturbing embedded planet) constitutes a linear perturbation to its background state. The 2D variables are given by weighted vertical averages of the (adiabatic) pressure and velocity perturbations to the locally isothermal background state of a 3D disc. Specifically, if we define
\begin{gather}
    \bar{\Sigma} \equiv \Sigma_p + \frac{P'}{c_\gamma^2},\\
    \bar{\bf{v}} \equiv \bar{v}_x{\bf{e}}_x +  \left(-\tfrac{3}{2}\Omega_p x + \bar{v}_y\right){\bf{e}}_y,
\end{gather}
then we see that our vertically averaged flow equations (\ref{2DMC}), (\ref{2Dxmom}) and (\ref{2Dymom}) are linearisations of the familiar 2D barotropic model,
\begin{subequations}\label{full2DEulereq}
\begin{align}
&{\bar{\bf{v}}}\cdot \nabla \bar{{\bf{v}}}  + 2{\bf{\Omega}}_p \times \bar{{\bf{v}}} = - \frac{1}{\bar{\Sigma}} \nabla \bar{P} - \nabla \bar{\Phi}_t - \nabla \Phi_p,\label{Full2Da}\\
&\nabla \cdot \left(\bar{\Sigma} \bar{\bf{v}}\right) = 0,\label{Full2Db}\\
&\bar{P} = K \bar{\Sigma}^\gamma,\label{Full2Dc}
\end{align}
\end{subequations}
where $\bar{\Phi}_t = -\tfrac{3}{2}\Omega_p^2 x^2$, and  $K = c_p^2\bar{\Sigma}_p^{1-\gamma}$ uniformly, so that $\dd \bar{P} = c_\gamma^2 \dd \bar{\Sigma}$. We remind the reader of the formal definitions of $P'$, $\bar{v}_x$ and $\bar{v}_y$, given in equation (\ref{eqnavg}), and our convention $c_\gamma^2 = \gamma c_p^2$. (Note we may treat (\ref{Full2Dc}) as an exact definition of $\bar{P}$.)

We note that the error in the system (\ref{full2DEulereq}) scales as the size of the quadratic non-linear terms, that is, $(q/h_\gamma^3)^2$. This is notably far smaller than the error introduced by any alternative averaging process, which would be of order $q/h_\gamma^3$. Recall we assumed $q \ll h_\gamma^3$ so that our 3D flow was linear, and that the errors introduced in the linearisation of the 3D system also scaled as $(q/h_\gamma^3)^2$. Furthermore, whilst we've addressed here only a local model of disc dynamics, this approach may be generalised to a global disc. It's sufficient for the background disc to satisfy a locally isothermal equation of state.

It's worth mentioning that the `surface density' $\bar{\Sigma}$ in the 2D model must really be thought of in terms of the vertically averaged pressure perturbation. Moreover, equation (\ref{Full2Db}) doesn't technically enforce physical mass conservation; mass conservation would be derived from a direct (unweighted) vertical integral of the 3D mass conservation equation (\ref{eqnEPd}). Instead, (\ref{Full2Db}) describes the evolution of the pressure due to a combination of compressive and buoyant motions; however, it takes precisely the same form as a 2D mass conservation equation (and for the remainder of this paper this is how we shall think of it).

In contrast, the 2D model's flow velocities ($\bar{v}_x$ and $\bar{v}_y$) can be derived from simple weighted vertical averages of their 3D counterparts. Finally, we remind the reader that the `planetary potential', $\Phi_p$, which forces this 2D system should be taken to be the vertical average of the 3D potential specified in (\ref{Phidefn}). This prescription is generally applicable to 2D disc models, and is discussed in more detail in section \ref{s:softening}.

\subsection{Stream function and Bernoulli invariant}\label{s:sfB}

The stream function and streamlines of our 2D vertically averaged flow are of particular importance near to corotation. This 2D behaviour is expected to contribute dominantly to the torque exerted on the planet, and for low-mass planets, the corotation torque is expected to scale with the fourth power of the horseshoe region width. To find and compute these streamlines, it's instructive to consider the exact 2D flow which solves (\ref{full2DEulereq}) (which is well approximated by our averaged flow), and exactly conserves the Bernoulli invariant (which we demonstrate below). Equation (\ref{Full2Db}) implies the existence of a stream function $\psi$. We take the definition
\begin{equation}
    \bar{\Sigma}\bar{\bf{v}} = -{\bf{e}}_z \times \nabla \psi.
\end{equation}
We may rewrite (\ref{Full2Da}) as
\begin{equation}\label{eqnBCd1}
\left(2 {\bf{\Omega}}_p + \nabla \times \bar{\bf{v}}\right) \times \bar{\bf{v}} + \nabla \left(\frac{1}{2}|\bar{\bf{v}}|^2 + \bar{W} + \bar{\Phi}_t + \Phi_p\right) = {\bf{0}},
\end{equation}
where $\dd \bar{W} = \dd \bar{P}/\bar{\Sigma}$. We define the Bernoulli invariant, $B$, as
\begin{equation}
    B = \frac{1}{2}|\bar{\bf{v}}|^2 + \bar{W} + \bar{\Phi}_t + \Phi_p,
\end{equation}
and introduce the potential vorticity (PV) (or vortensity)
\begin{equation}\label{zetadef}
    \zeta = \frac{2 \Omega_p + {\bf{e}}_z\cdot (\nabla \times \bar{\bf{v}})}{\bar{\Sigma}}.
\end{equation}
We may then rewrite equation (\ref{eqnBCd1}) as
\begin{gather}
\begin{split}
\bar{\Sigma}\zeta {\bf{e}}_z \times \bar{\bf{v}} + \nabla B = 0\\
\implies - \zeta {\bf{e}}_z \times\left({\bf{e}}_z \times \nabla \psi\right) + \nabla B = 0\\
\implies \nabla B = -\zeta\nabla \psi.
\end{split}
\end{gather}
That is to say, the Bernoulli function is constant on streamlines, $B = B(\psi)$, and further the PV,
\begin{equation}\label{eqnPVC1}
    - \frac{\dd B}{\dd \psi} = \zeta(\psi)
\end{equation}
is also conserved. We now further impose that $\zeta$ is uniformly constant upstream (which holds to leading order within the local approximation), so that
\begin{equation}
    \zeta \equiv \frac{\Omega_p}{2 \Sigma_p}.
\end{equation}
Combining this with the definition (\ref{zetadef}) implies that linearly
\begin{equation}
    \upartial_x \bar{v}_y - \upartial_y \bar{v}_x - \frac{\Omega_p}{2 c_\gamma^2} \frac{P'}{\Sigma_p} = 0.
\end{equation}
In other words, the PV is uniform. Integrating equation (\ref{eqnPVC1}) (and setting the arbitrary constant of integration to zero) then gives an expression for the stream function for the 2D flow
\begin{equation}
    -\frac{\Omega_p}{2 \Sigma_p}\psi = \frac{1}{2}|\bar{\bf{v}}|^2 + \bar{W} + \bar{\Phi}_t + \Phi_p.
\end{equation}
That is to say, our 2D vertically averaged flow has streamlines which are the contours of the stream function
\begin{equation}\label{2Dpsi}
    \frac{\psi}{\Sigma_p} = \frac{3}{4}\Omega_p x^2 - \frac{2}{\Omega_p}\left(\frac{P'}{\Sigma_p}+\Phi_p\right) + 3 x \bar{v}_y + \mathcal{O}\left(\frac{q^2}{h_\gamma^6}\frac{c_\gamma^2}{\Omega_p}\right).
\end{equation}

\subsection{On the corotation singularity}\label{s:crtsing}

Importantly, (\ref{2Dpsi}) allows us to write down the equation for the radial displacement of fluid elements, $\bar{\xi}_x$, which we take to satisfy
\begin{equation}\label{xidefn}
    \bar{\bf{v}}\cdot \nabla \bar{\xi}_x = \bar{v}_x,
\end{equation}
with $\bar{\xi}_x = 0$ far upstream. If a given streamline (or contour of $\psi$) has radial location $x_0$ far upstream, we have that
\begin{equation}
    \bar{\xi}_x = x - x_0.
\end{equation}
Comparing values of $\psi$ far upstream to the point $(x,y)$, we see that at leading order
\begin{multline}
    \psi = \frac{3}{4}\Sigma_p\Omega_p x_0^2 = \frac{3}{4}\Sigma_p\Omega_p \left(\bar{\xi}_x-x\right)^2 \\ = \frac{3}{4}\Sigma_p\Omega_p x^2 - \frac{2\Sigma_p}{\Omega_p}\left(\frac{P'}{\Sigma_p}+\Phi_p\right) + 3 x \Sigma_p\bar{v}_y,
\end{multline}
\begin{equation}\label{eqnxidef}
\implies {\bar{\xi}_x}^2 - 2 x \bar{\xi}_x + \frac{8}{3\Omega_p^2}\left(\frac{P'}{\Sigma_p}+\Phi_p\right) - \frac{4 x \bar{v}_y}{\Omega_p} = 0,
\end{equation}
\begin{equation}\label{eqnxi}
\implies \bar{\xi}_x = x \pm\sqrt{x^2 -\frac{8}{3\Omega_p^2}\left(\frac{P'}{\Sigma_p}+\Phi_p\right) + \frac{4 x \bar{v}_y}{\Omega_p}},
\end{equation}
where the choice of $+$ or $-$ is determined by whether the fluid element has just undertaken a horseshoe turn or not. Here, $P'$ and $\bar{v}_y$ may be taken to be the linear solutions to the 2D equations of motion, which we reduce to independent second order equations in section \ref{s:reduction}.

Equation (\ref{eqnxidef}) captures the essence of the corotation singularity, and what is meant by the `non-linear' corotation torque. The linear theory such as that applied in \citet{goldreich_excitation_1979} and \citet{tanaka_threedimensional_2002} may be thought to implicitly discard the quadratic term ${\bar{\xi}_x}^2$, which is non-negligible for $x = \mathcal{O}\left(\sqrt{\frac{q}{h^3}}H\right)$. In this way, the linear equation for the particle displacement experiences a singularity at $x = 0$, which must be resolved non-linearly (though there exist linear equations which remain valid at corotation for $\bar{v}_x$, $\bar{v}_y$ and $P'$). The singularity is introduced in linear analyses when the linearized azimuthal momentum equation, in our case equation (\ref{2Dymom}), is used to directly re-arrange for $\bar{v}_y$, namely by integrating with respect to azimuthal coordinate $y$ or $\theta$ (which may look like dividing by $im(\Omega - \Omega_p)$ in Fourier space). More specifically, this `first integral' equation for $\bar{v}_y$ must become non-linear to remain valid near to corotation, just as the azimuthal integral of $\bar{v}_x$ becomes large enough that inclusion of the quadratic term $\bar{\xi}_x^2$ in equation (\ref{eqnxidef}) is necessary.

Importantly, this means that any materially conserved quantity $Q(\psi) = Q(\bar{\xi}_x - x)$, will experience the same singularity as the radial displacement, which appears when the quadratic term in (\ref{eqnxidef}) is neglected. A gradient of (the materially conserved) potential vorticity over the coorbital region will therefore induce a singularity in the linear equations of motion. In order to resolve this singularity, the steady distribution of PV, which is advected by the flow induced by the planet within the horseshoe region, must be ascertained. In this sense, the corotation torque is `non-linear'. The corotation torque that arises depends very strongly on the geometry of the base flow, demanding an accurate model for the flow induced by a planet. We find this flow in the case of a low-mass planet in this paper.

\subsection{2D mode orthogonality and torque decomposition}\label{s:td}

For $\gamma < 2$, the unforced linearized equations of motion in a disc admit a `2D mode' solution with $v'_z = 0$. This 2D mode is a member of a wider family of modes, whose axisymmetric members are discussed in \citet{lubow_wave_1993}. Here we demonstrate that the 2D mode is orthogonal to the remaining set of (non-axisymmetric) 3D disc modes, and show how the torque may be decomposed into 2D and 3D components. For $\gamma > 2$, a different family of modes exists which does not include such a mode with $v'_z = 0$. Indeed, imposing $v'_z = 0$ in this case yields an unnormalisable mode of infinite energy. Appropriately normalised, the 2D mode may be written as
\begin{subequations}
\begin{align}\label{2dzprof}
    &v'_{x,0} = \bar{v}_x(x,y)\sqrt{2-\gamma}\exp{\left(\tfrac{(\gamma-1)z^2}{2 H_\gamma^2}\right)},\\
    &v'_{y,0} = \bar{v}_y(x,y)\sqrt{2-\gamma}\exp{\left(\tfrac{(\gamma-1)z^2}{2 H_\gamma^2}\right)},\\
    &p'_0 = P'(x,y)\sqrt{2-\gamma}\frac{\rho_p(z)}{\Sigma_p}\exp{\left(\tfrac{(\gamma-1)z^2}{2 H_\gamma^2}\right)},\\
    &v'_{z,0} = 0, \quad \rho'_0 = p'_0/c_\gamma^2,
\end{align}
\end{subequations}
where again
\begin{equation}\label{eqnavg2}
    \bar{v}_x \equiv \left\langle v'_x\right\rangle, \quad \bar{v}_y \equiv \left\langle v'_y\right\rangle, \quad P' \equiv \Sigma_p \left\langle p'/\rho_p(z)\right\rangle.
\end{equation}
We may then decompose, for example, the radial velocity into its 2D mode component and an orthogonal complement, which we call $v'_{x,c}$, describing the rest of the disc's 3D motions, which include for example inertial and gravity waves. 
\begin{equation}
    v'_x = v'_{x,0} + v'_{x,c}.
\end{equation}
Now, $v'_{x,0}$ and $v'_{x,c}$ are everywhere orthogonal with respect to the inner product involving the density
\begin{equation}
    \left\langle f,g \right\rangle = \frac{1}{\Sigma_p}\int_{-\infty}^{\infty} \rho_p(z) f(z) g(z) \,\dd z.
\end{equation}
This may be seen from the relation for the vertical average $\left\langle\,\cdots\right\rangle$ in terms of the density-weighted inner product
\begin{equation}
    \left\langle\,\cdots\right\rangle \equiv \left\langle\,\cdots,\frac{1}{\sqrt{\gamma}}\exp{\left(\tfrac{(\gamma-1)z^2}{2 H_\gamma^2}\right)}\right\rangle.
\end{equation}
Specifically, the orthogonality follows via
\begin{align*}
    \left\langle v'_{x,0}, v'_{x,c} \right\rangle &= \left\langle v'_{x,0}, v'_{x} - v'_{x,0} \right\rangle \\
    &= \sqrt{2-\gamma}\,\bar{v}_x\left\langle \exp{\left(\tfrac{(\gamma-1)z^2}{2 H_\gamma^2}\right)}, v'_{x}\right\rangle \\& \qquad - (2-\gamma)\bar{v}_x^2\left\langle\exp{\left(\tfrac{(\gamma-1)z^2}{2 H_\gamma^2}\right)},\exp{\left(\tfrac{(\gamma-1)z^2}{2 H_\gamma^2}\right)}\right\rangle \\
    &= \sqrt{\gamma(2-\gamma)}\,\bar{v}_x\left\langle v'_x\right\rangle - (2-\gamma)\bar{v}_x^2\sqrt{\frac{\gamma}{2-\gamma}}\\
    &=0.
\end{align*}
As a result, we obtain the Parseval identity,
\begin{equation}
\left\langle v'_x, v'_y \right\rangle = \left\langle v'_{x,0}, v'_{y,0} \right\rangle + \left\langle v'_{x,c}, v'_{y,c}\right\rangle,
\end{equation}
as well as equivalent identities for any two variables drawn from the set $\{v'_x, v'_y, p'/\rho_p(z)\}$. In particular, this allows us to decompose the radial angular momentum flux, $F_A$, into independent components associated with the 2D mode and 3D remainder, since
\begin{multline}\label{fluxdecomp}
    F_A \approx r_p\iint \rho_p(z) v'_x v'_y \,\dd z\,\dd y \\ = r_p\iint \rho_p(z) v'_{x,0} v'_{y,0} \,\dd z\,\dd y + r_p\iint \rho_p(z) v'_{x,c} v'_{y,c} \,\dd z \,\dd y \\ = \sqrt{\gamma(2 - \gamma)}r_p \Sigma_p\int_{-\infty}^{\infty} \bar{v}_x\bar{v}_y \,\dd y + F_A^{\text{3D}} \equiv F_A^{\text{2D}} + F_A^{\text{3D}}
\end{multline}
It's in this sense that the torque on the planet for a general adiabatic index $\gamma$ may be separated into 2D and 3D contributions. The factor of $\sqrt{\gamma(2-\gamma)}$ (approximately 92\% for $\gamma = 1.4$) may be shown to be the fraction of torque imparted into the 2D mode by an external potential (assumed to be $z$-independent) at a Lindblad resonance of order $m \ll 1/h_\gamma$ (see for example equation (45) in \citet{lubow_threedimensional_1998} and appendix B1 of \citet{bate_excitation_2002}). In this way, naïvely approximating the 3D flow velocities by their averaged values inadvertently recovers the total flux in this simplified scenario, as this approximation removes the factor of $\sqrt{\gamma(2-\gamma)}$ from the flux expression. This follows from the applicability of resonant torque excitation theory to this regime.

This principle is at play in figure 2 of \citet{zhu_planet-disk_2012}. The top left panel depicts the simulated torque density excited in 3D discs which admit adiabatic perturbations to an isothermal background state. Notably, for $x\gtrsim2H_\gamma$, and for each value of $\gamma$ considered, \citet{zhu_planet-disk_2012} recover torque densities matching the case $\gamma = 1$ (in which the 2D mode is $z$-independent and contributes almost all of the torque). The torque density in this `outer' region scales as $x^{-4}$ \citep{goldreich_disk-satellite_1980}, so that their $\gamma$-dependent scaling of the $x$-axis with $H_\gamma$ and $y$-axis with $c_\gamma^4$ does not meaningfully affect the outer torque density curves. This torque density agreement is achieved despite a considerable fraction of the flux being carried by gravity waves and inertial waves when $\gamma > 1$. That is, the torque density for $x\gtrsim2H_\gamma$ matches that obtained via a naïve approximation of the 3D flow by its vertical average, as defined in (\ref{eqnavg}).

There is however in the planet-disc interaction problem an important (indeed, a dominant) contribution from azimuthal modes with $m \sim 1/h_\gamma$. The vertical profile of the planetary potential plays a key role here too. Indeed, inertial waves and gravity waves excited near the planet also impact the torque exerted on the planet significantly, contributing to $F_A^{\text{3D}}$. The problem of resolving the spectrum of gravity waves excited by an embedded planet and the flux they carry is very challenging, and has received only limited attention \citep{zhu_planet-disk_2012,lubow_analytic_2014,mcnally_low-mass_2020,bae_observational_2021,ziampras_buoyancy_2023}. Numerical approaches face difficulties resolving the complex and fine structure of the waves, and quantitative analytical approaches struggle since the gravity waves are not separable in the vertical direction.

\section{Reduction to independent second order linear equations}
\label{s:reduction}

In section \ref{s:governingeqs}, we showed how the 3D equations governing the flow near a low-mass planet may be manipulated into the same form as the familiar 2D local equations for mass, PV and momentum conservation
\begin{subequations}\label{2DEuler}
\begin{align}
&D P' + \Sigma_p c_\gamma^2\left(\upartial_x \bar{v}_x + \upartial_y \bar{v}_y\right) = 0,\label{2DEulerMC}\\
&\upartial_x \bar{v}_y - \upartial_y \bar{v}_x - \frac{\Omega_p}{2 c_\gamma^2} \frac{P'}{\Sigma_p} = 0,\label{2DEulera}\\
&D\bar{v}_x - 2\Omega_p \bar{v}_y + \frac{1}{\Sigma_p}\upartial_x P' = -\upartial_x \Phi_p,\label{2DEulerb}\\
&D\bar{v}_y + \frac{1}{2}\Omega_p \bar{v}_x + \frac{1}{\Sigma_p}\upartial_y P' = -\upartial_y \Phi_p,\label{2DEulerc}
\end{align}
\end{subequations}
where $D \equiv -\frac{3}{2}x\Omega_p\upartial_y$ is the leading order advective operator arising from the background shear flow. We now derive with no further approximations simple, second order linear equations from these which describe both the spiral wave excitation and horseshoe streamlines. It's helpful to non-dimensionalise the equations. We first replace
\begin{equation}
    x \to H_\gamma x, \quad y \to H_\gamma y,
\end{equation}
so that $x$ and $y$ now measure distances in units of `adiabatic scale heights'. We also let
\begin{equation}
    \mathcal{D} = -\frac{3}{2}x\upartial_{y},
\end{equation}
and introduce the non-dimensional potential $\hat{\phi}_p$
\begin{multline}
    \Phi_p(x,y) = \frac{q}{h_\gamma^3}c_\gamma^2\hat{\phi}_p\left(s\right), \quad s = \sqrt{x^2+y^2},\\ \implies \hat{\phi}_p(s) = -\frac{\ee^{\frac{1}{4}s^2}}{\sqrt{2\upi}}K_0\left(\tfrac{1}{4}s^2\right) \sim -\frac{1}{s} + \mathcal{O}\left(\frac{1}{s^3}\right).
\end{multline}
We define non-dimensional, scaled $x$- and $y$- velocity and `enthalpy' perturbations $u(x,y)$, $v(x,y)$ and $W(x,y)$ via
\begin{equation}\label{nondim}
    \bar{v}_x \equiv \frac{q}{h_\gamma^3}c_\gamma u, \quad \bar{v}_y \equiv \frac{q}{h_\gamma^3}c_\gamma v, \quad P' \equiv \frac{q}{h_\gamma^3}c_\gamma^2\Sigma_p W.
\end{equation}
Equations (\ref{2DEulera}), (\ref{2DEulerb}) and (\ref{2DEulerc}) become
\begin{subequations}\label{2DNDeqs}
\begin{align}
&\upartial_x v - \upartial_y u - \frac{1}{2}\chi = - \frac{1}{2}\hat{\phi}_p,\label{2DNDeqsa}\\
&\mathcal{D} u - 2v + \upartial_x \chi = 0,\label{2DNDeqsb}\\
&\mathcal{D} v + \frac{1}{2}u + \upartial_y \chi = 0,\label{2DNDeqc}
\end{align}
\end{subequations}
where $\chi \equiv W + \hat{\phi}_p$. Additionally, `mass conservation', (\ref{2DEulerMC}) becomes
\begin{equation}
\mathcal{D}\chi + \upartial_x u + \upartial_y v = \mathcal{D}\hat{\phi}_p,
\end{equation}
which may also be derived from (\ref{2DNDeqsa}), (\ref{2DNDeqsb}) and (\ref{2DNDeqc}). It's clear at this stage that the flow is indeed linear so long as $q \ll h_\gamma^3$.

Remarkably, these equations are independent of $\gamma$! They are analogous to the equations considered by \citet{paardekooper_width_2009}; however, it's worth noting that the approximations adopted in their subsequent analysis led to three sources of inaccuracy. Firstly, the use of decaying boundary conditions instead of a radiation condition, as well as the use of a softened planetary potential both introduce discrepancies. Additionally, motivated by its validity in the case of test particle dynamics, they further approximated the above equations near to corotation by setting $\mathcal{D} = 0$ (before then differentiating with respect to $x$). The resulting system omits large factors at corotation in the fluid case. Note that $\upartial_x\mathcal{D} \to -\tfrac{3}{2}\upartial_y$ as $x \to 0$, which does not vanish at corotation.

Before continuing the derivation, it's helpful also to note the equation for PV conservation in differential form,
\begin{equation}\label{eqPVdiff}
\mathcal{D}(\upartial_x v - \upartial_y u) + \frac{1}{2}(\upartial_x u + \upartial_y v) = 0.
\end{equation}
To proceed, we take $\mathcal{D}$(\ref{2DNDeqsb}) and combine with (\ref{2DNDeqc}), and then take $\mathcal{D}$(\ref{2DNDeqc}) and combine with (\ref{2DNDeqsb}). This yields
\begin{subequations}
\begin{align}
&(\mathcal{D}^2 + 1)u = - \mathcal{D} \upartial_x \chi - 2 \upartial_y \chi,\label{eqvx02}\\
&(\mathcal{D}^2 + 1)v = - \mathcal{D} \upartial_y \chi + \frac{1}{2} \upartial_x \chi.\label{eqvy02}
\end{align}
\end{subequations}
Now, it may be shown from equations (\ref{2DNDeqsa}) and (\ref{eqPVdiff}) that
\begin{equation}
\mathcal{D} \upartial_x \chi + 2 \upartial_y \chi = \mathcal{D} \upartial_x \hat{\phi}_p + 2 \upartial_y \hat{\phi}_p - \nabla^2 u + 3\upartial_y(\chi - \hat{\phi}_p),
\end{equation}
and
\begin{equation}
\mathcal{D} \upartial_y \chi - \frac{1}{2} \upartial_x \chi = \mathcal{D} \upartial_y \hat{\phi}_p - \frac{1}{2} \upartial_x \hat{\phi}_p - \nabla^2 v.
\end{equation}
Equations (\ref{eqvx02}) and (\ref{eqvy02}) become
\begin{subequations}
\begin{align}
&(\mathcal{D}^2 + 1 - \nabla^2)u + 3 \upartial_y \chi = \upartial_y \hat{\phi}_p - \mathcal{D} \upartial_x \hat{\phi}_p,\label{eqvx03}\\
&(\mathcal{D}^2 + 1 - \nabla^2)v = - \mathcal{D} \upartial_y \hat{\phi}_p + \frac{1}{2} \upartial_x \hat{\phi}_p.\label{eqvy03}
\end{align}
\end{subequations}
Now, to find an equivalent equation for $\chi$, we take $\mathcal{D}^2$(\ref{2DNDeqsa}). We're then able to use equation (\ref{eqPVdiff}), the momentum equations (\ref{2DNDeqsb}) \& (\ref{2DNDeqc}) and the PV equation (\ref{2DNDeqsa}) again to deduce
\begin{equation}\label{eqPV03}
(\mathcal{D}^2 + 1 - \nabla^2)\chi + 3\upartial_y u = (\mathcal{D}^2 + 1)\hat{\phi}_p.
\end{equation}
We now define the linearized Riemann invariants\footnote{The interpretation of $J_\pm$ as linearized Riemann invariants is clarified in section \ref{s:waveevo}.} $J_\pm$:
\begin{equation}
    J_\pm \equiv u \pm \chi.
\end{equation}
Combining equations (\ref{eqPV03}) and (\ref{eqvx03}), it follows that
\begin{subequations}
\begin{align}
&\left[\mathcal{D}^2 + 1 \pm 3 \upartial_y - \nabla^2\right]J_{\pm} =  \left[\upartial_y - \mathcal{D} \upartial_x \pm (\mathcal{D}^2 + 1)\right]\hat{\phi}_p,\label{eqJpm0F}\\
&\left[\mathcal{D}^2 + 1 - \nabla^2\right]v =  - \mathcal{D} \upartial_y \hat{\phi}_p + \frac{1}{2}\upartial_x \hat{\phi}_p.\label{eqvy0F}
\end{align}
\end{subequations}
Now, equation (\ref{eqvy0F}) is the parabolic cylinder equation\footnote{More specifically, it becomes the parabolic cylinder equation following an azimuthal mode decomposition or Fourier transform.} derived by \citet{artymowicz_wave_1993} for the azimuthal velocity perturbation, but equation (\ref{eqJpm0F}) is novel. Importantly, it provides a simple and non-singular description of the behaviour of the radial velocity and enthalpy at corotation. It describes the linear excitation of the profiles of the Riemann invariants of \citet{goodman_planetary_2001}, which are conserved non-linearly further from the planet. These equations also accurately capture the behaviour of the coorbital flow, including the horseshoe dynamics.

It is worth noting and crediting the similar equations derived by \citet{heinemann_excitation_2009}, who studied the excitation of density waves via turbulence. In the \hyperref[appx]{appendix}, we discuss the numerical solution of equations (\ref{eqJpm0F}) and (\ref{eqvy0F}) to high accuracy, and we present these solutions in the next section.

The inner limit of equations (\ref{eqJpm0F}) and (\ref{eqvy0F}) (that is, the simplified equations valid close to corotation taking $x \ll H$) may be written as
\begin{subequations}
\begin{align}
&\left[1 \pm 3 \upartial_y - \nabla^2\right]J_{\pm} =  \left[\upartial_y - \mathcal{D} \upartial_x \pm (\mathcal{D}^2 + 1)\right]\hat{\phi}_p,\label{eqJpm0Fi}\\
&\left[1 - \nabla^2\right]v =  - \mathcal{D} \upartial_y \hat{\phi}_p + \frac{1}{2}\upartial_x \hat{\phi}_p.\label{eqvy0Fi}
\end{align}
\end{subequations}
Note that terms including $\mathcal{D}$ acting on $\hat{\phi}_p$ may not be neglected, as in this case $\hat{\phi}_p$ has a (logarithmic) singularity at the origin, with the effect that such terms are non-negligible. For comparison, the equivalent equation from \citet{paardekooper_width_2009} expressed in our notation reads
\begin{equation}\label{eqPP09b}
    \left[1 - \partial_x^2 - 4\partial_y^2\right]\chi = \phi_p.
\end{equation}

\subsection{Connection with resonant wave excitation theory}\label{s:LR}

One may deduce directly from (\ref{eqvy0F}) the locations of the `effective' Lindblad resonances. In a pressureless disc composed of test-particles, the Lindblad resonances are located at orbital radii such that the epicyclic frequency, $\kappa(r)$, is an integer multiple of the interaction rate, $\Omega - \Omega_p$, of the test particles with the planet. In this way, successive interactions of a test particle with the planet will administer in phase `kicks' to the test particle. In the 2D gas-dynamic case, inertio-acoustic wave disturbances instead oscillate with squared frequency $\omega^2 = \kappa^2 + c^2 \left|{\bf{k}}\right|^2$ for wavevector ${\bf{k}}$ and sound speed $c$. In this way, disturbances of large wavenumber have a larger frequency, which has the effect that the Lindblad resonances of large order need not have a very small interaction rate. In this way, the Lindblad resonances all stand off and pile up a finite distance from the planet's orbital radius, namely at $x = \pm\tfrac{2}{3}H_\gamma$.

Furthermore, in the gas-dynamic case the contribution of the radial wavenumber $k_x$ to the wave frequency $\omega$ leads to the spreading out of each Lindblad resonance radially.

The resonance locations may equivalently be defined in the gas-dynamic context as where the solution for a given azimuthal mode changes from evanescent to wavelike. It's near this turning point, where the wave has zero frequency (and correspondingly the most time to be excited), that the forcing has most influence on the wave excitation. We note that the operator $\mathcal{D}^2 + 1 - \nabla^2$ is hyperbolic in the wave-permitting region $\left|x\right| > \tfrac{2}{3}H_\gamma$, and elliptic in the region where all modes are evanescent, namely $\left|x\right| < \tfrac{2}{3}H_\gamma$. This signposts that the Lindblad resonances must all stand off a distance $\tfrac{2}{3}H_\gamma$ from the planet's orbital radius. More specifically, for the azimuthal disc mode proportional to $\ee^{\ii m \theta}$, recalling the original definition $y = r_p\theta$ and reintroducing dimensions, (\ref{eqvy0F}) becomes
\begin{equation}
    \left(- \frac{9}{4} \Omega_p^2 \frac{x^2}{r_p^2}m^2 + \Omega_p^2 + c_\gamma^2\frac{m^2}{r_p^2} - c_\gamma^2\upartial_x^2 \right)v = \cdots.
\end{equation}
We see that were it not for the acoustic terms in the above operator, namely $c_\gamma^2 m^2/r_p^2$ and $c_\gamma^2\upartial_x^2$, we would recover the positions of the close-in Lindblad resonances for test-particles by setting this operator to $0$, that is, $r_{L,\pm m} = r_p\left(1 \pm \tfrac{2}{3 m}\right)$. Appropriately, at these resonances, solving for $v$ would involve division by $0$. Including the acoustic terms yields a parabolic cylinder equation for $v_y$ whose solution is evanescent for $\left|x\right| < \tfrac{2}{3}H_\gamma\sqrt{1 + (h_\gamma m)^{-2}}$, and wavelike for $\left|x\right| > \tfrac{2}{3}H_\gamma\sqrt{1 + (h_\gamma m)^{-2}}$. The effective Lindblad resonances close to the planet (with $h_\gamma m \gtrsim 1$) therefore have locations
\begin{equation}
    r_{L,\pm m} = r_p \pm \frac{2}{3}H_\gamma\sqrt{1 + (h_\gamma m)^{-2}}.
\end{equation}

\section{Results}
\label{s:results}

In this section, numerical solutions to equations (\ref{eqJpm0F}) and (\ref{eqvy0F}) are presented and discussed. We compare our flow solution near to corotation with flows computed in 3D simulations, for example by \citet{lega_outwards_2015}, \citet{fung_3d_2015} and \citet{masset_horseshoe_2016}. We further compare the excited wave profiles with those found in 2D studies. We find good agreement with 3D one-sided Lindblad torque estimates, which are typically a factor of 2-3 lower than 2D values.

\begin{figure*}\centering
  \includegraphics[width=0.32\linewidth]{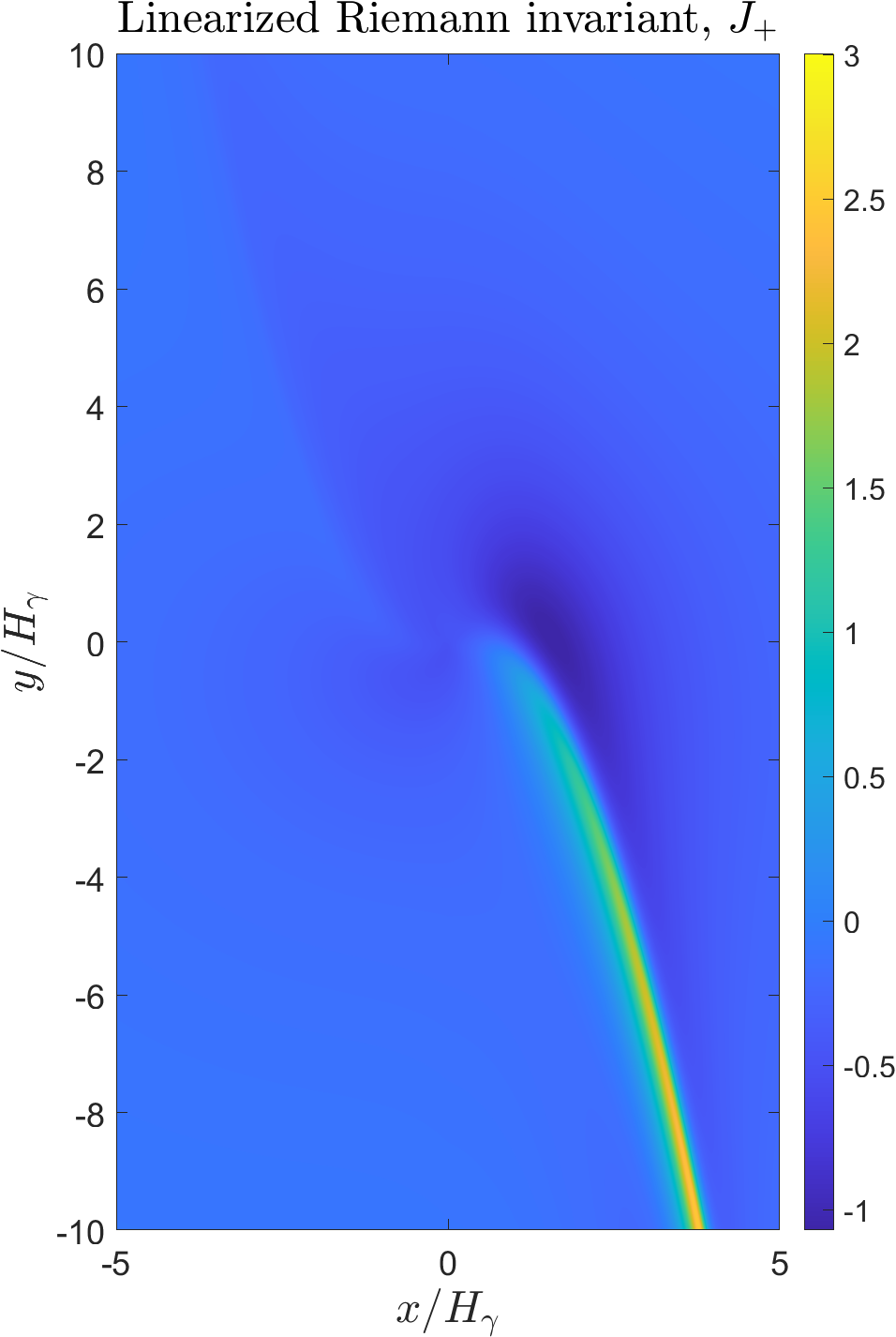}
  \includegraphics[width=0.32\linewidth]{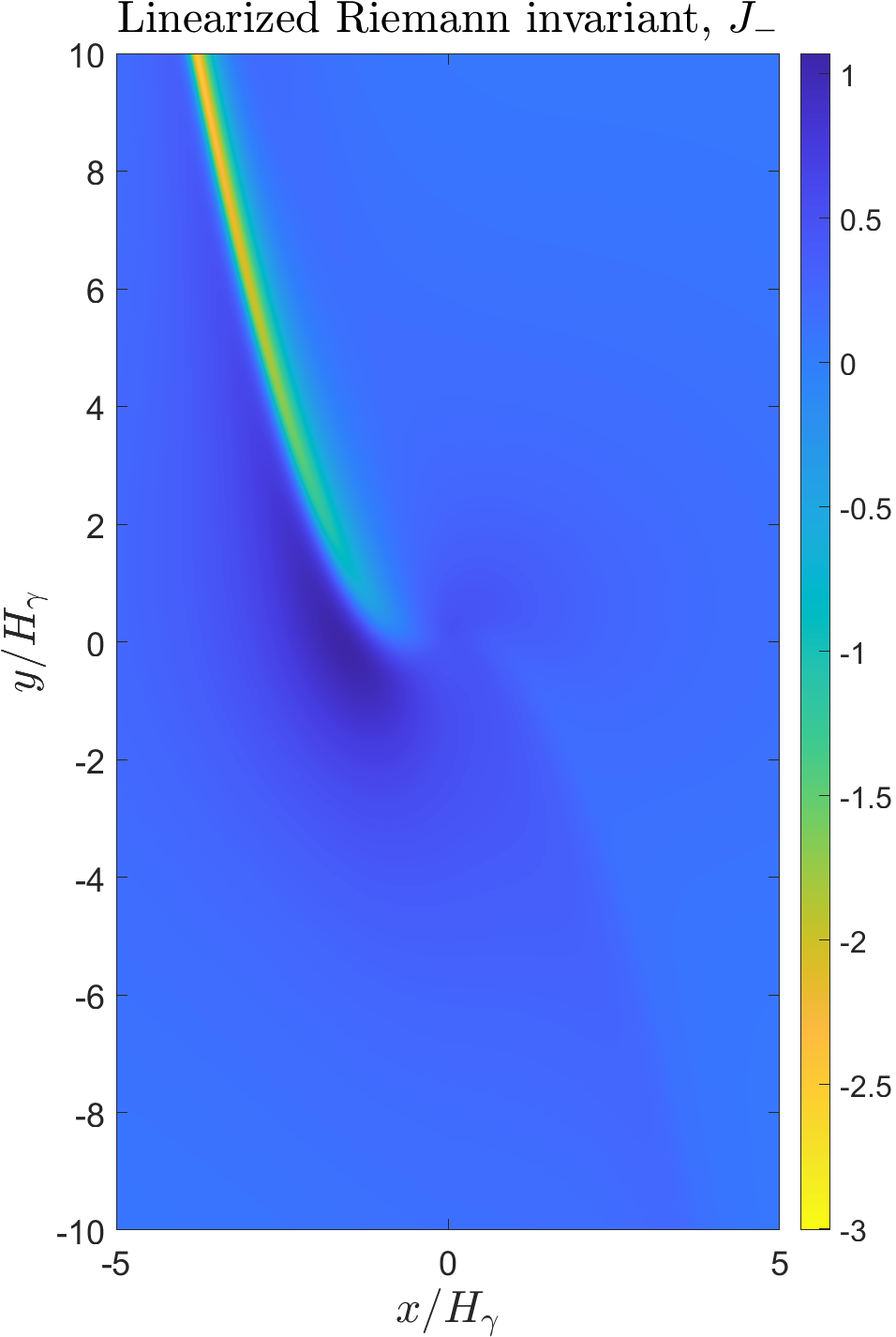}
  \includegraphics[width=0.32\linewidth]{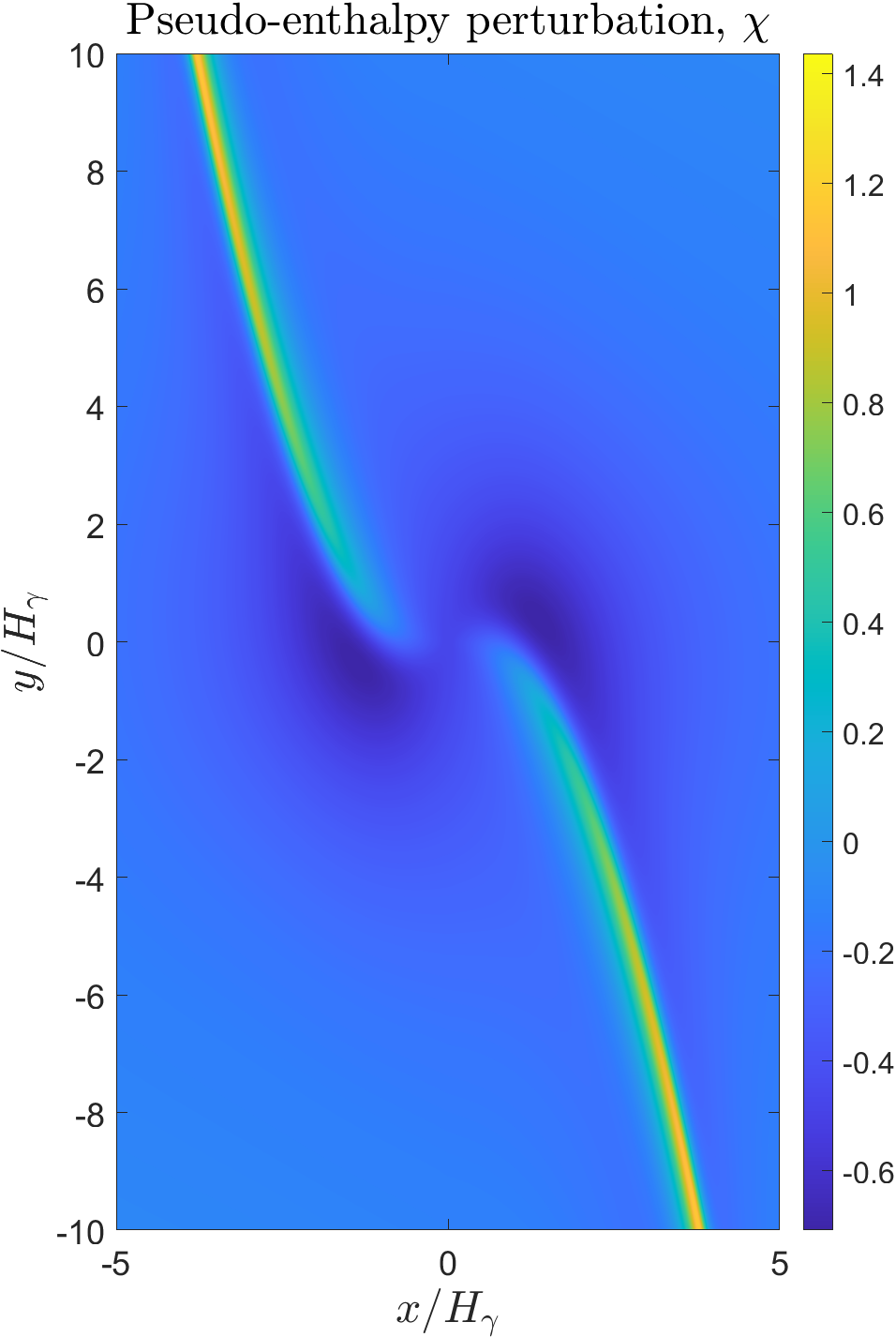}\vspace{1mm}
  \includegraphics[width=0.32\linewidth]{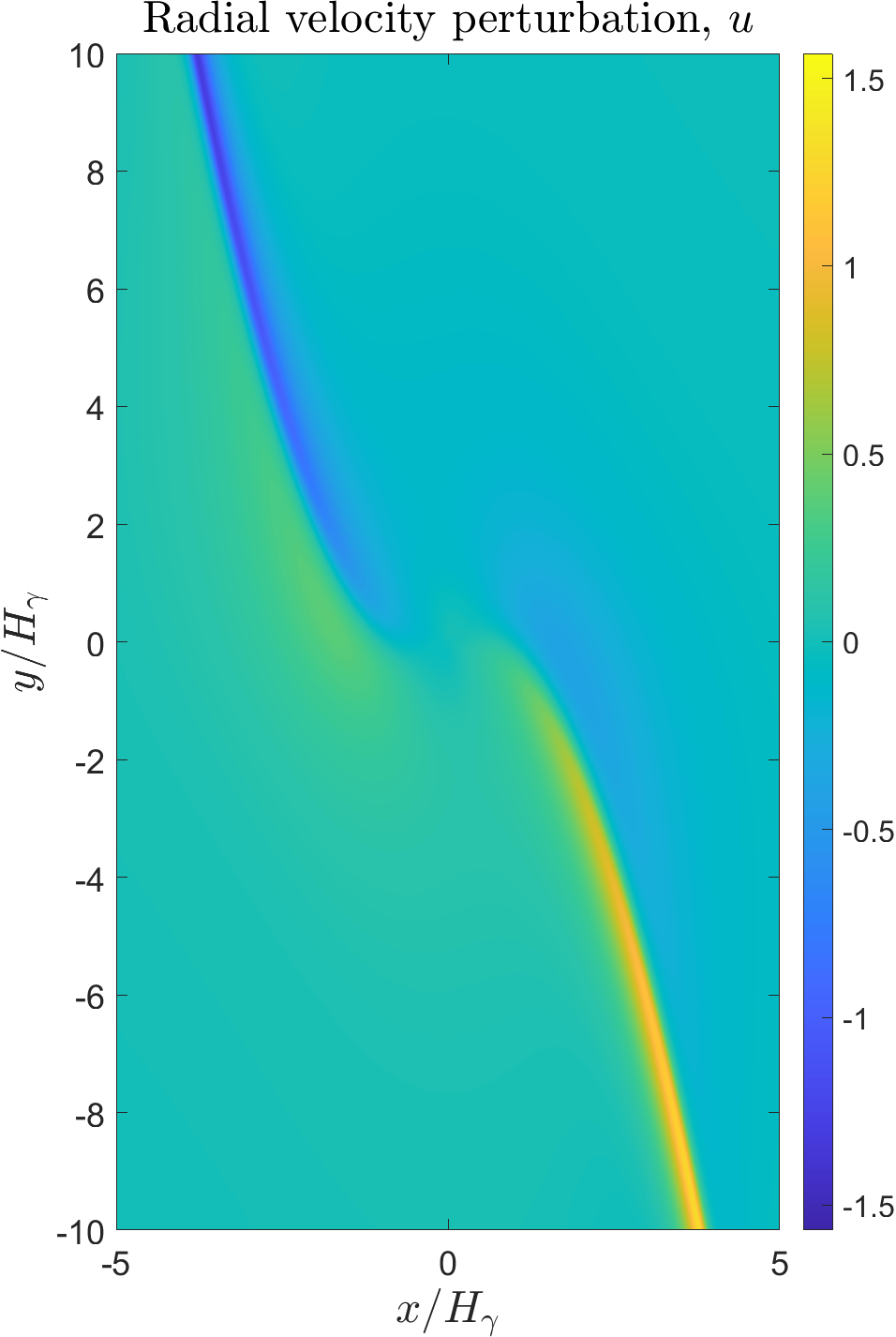}
  \includegraphics[width=0.32\linewidth]{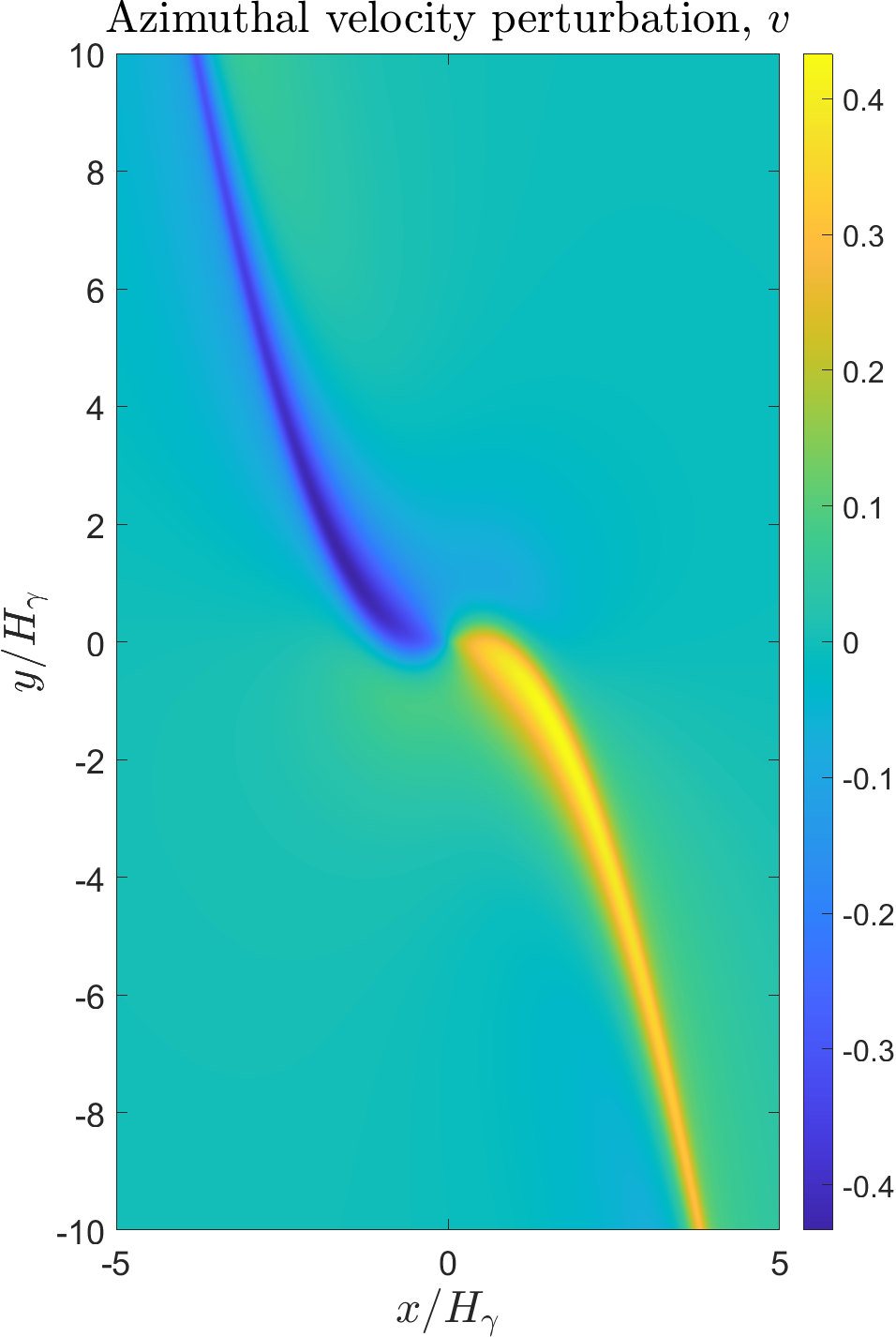}
  \includegraphics[width=0.32\linewidth]{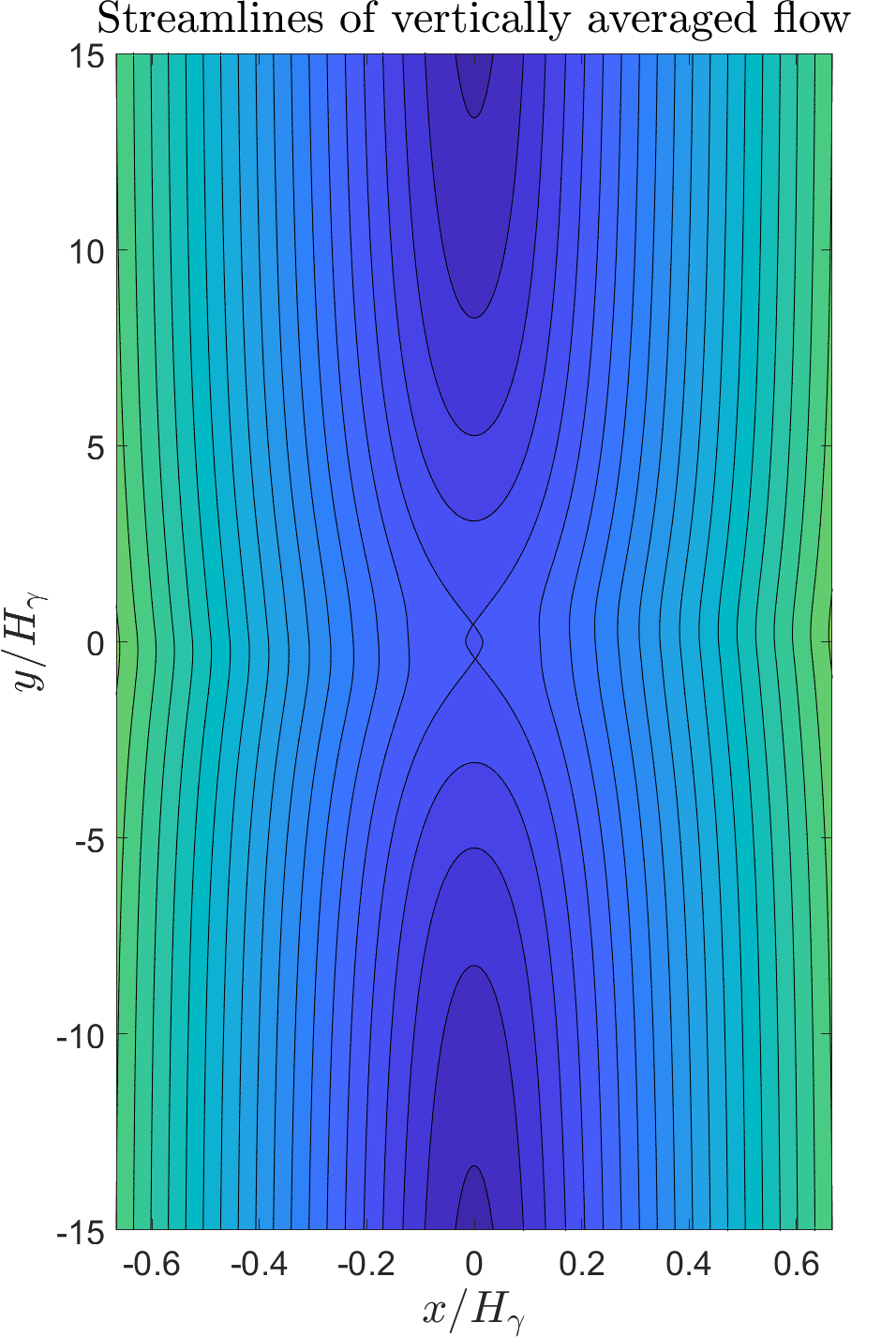}
  \caption{Graphs of non-dimensionalised linearized Riemann invariants $J_+$ and $J_-$ as well as pseudo-enthalpy $\chi = W + \hat{\phi}_p$, radial velocity $u$ and azimuthal velocity perturbation $v$ near the planet. The scalings for the variables are given in equation (\ref{nondim}). The planet is located at the centre of each panel. Data were obtained via solution of equations (\ref{eqJpm0F}) and (\ref{eqvy0F}) using the method described in the \hyperref[appx]{appendix}. The bottom right panel depicts streamlines of the 2D mode near the planet.}
  \label{3Disofigs1}
\end{figure*}
\subsection{Flow streamlines and the horseshoe width}\label{s:flowsl}

Real-space plots for the solutions of equations (\ref{eqJpm0F}) and (\ref{eqvy0F}) for each flow variable ($u$, $v$, $\chi$, $J_+$, $J_-$, as well as the stream function $\psi$), are shown in figure \ref{3Disofigs1}. The planet is located at the origin of each figure. We imposed radiation boundary conditions in the $x$-direction, specifying that our numerical solution includes no incoming waves, and used a Fourier transform method in $y$. Our numerical method is accurate and tailored specifically to this problem, and the plotted solutions have an uncertainty of $1 \times 10^{-5}$. Details on this numerical procedure are discussed in the \hyperref[appx]{appendix}.

Qualitatively, the vertically averaged flow field streamlines depicted in the bottom right panel of figure \ref{3Disofigs1} are comparable to those obtained with a `softened' potential with smoothing length $b = 0.4H_\gamma$. Namely, they agree in their prediction for the horseshoe width, and the streamlines are only weakly affected by the presence of the density waves. This is to be expected, as this particular choice for $b$ is known to match well the horseshoe width measured in 3D simulations \citep{lega_outwards_2015,masset_horseshoe_2016}.

In this case there is however a noticeable difference in the excited density wave's amplitude. We find via a second analogous calculation that the `softened' potential excites a wave with peak amplitude 30\% greater than that of the 2D mode in both the profiles of $J_+$ and $v$. This wave correspondingly transports an angular momentum flux inflated by 55\%, depicted in figure \ref{flux}. This numerical discrepancy highlights the well-known result that softening prescriptions are unable to simultaneously capture both the corotation torque as well as the Lindblad torque accurately \citep{masset_coorbital_2002}. Indeed, in figure 5 of \citet{tanaka_three-dimensional_2024} one observes that torque components vary by up to a factor of 3 as the smoothing length is varied between $0.3$ and $0.7$. 

The horseshoe region semi-thickness $x_s$ may be found from our solution by first noting that $\chi$, $u$ and $v$ decay rapidly to $0$ as $y \to \infty$ in the region $|x| < \frac{2}{3}H_\gamma$. The leading order far-field horseshoe streamlines therefore have reflectional symmetry about $x = 0$. We showed in section \ref{s:sfB} that the stream function in our vertically averaged flow may be expressed as
\begin{equation}\label{LOsf}
    \frac{\psi}{\Sigma_p} = \frac{3}{4}\Omega_p x^2 - \frac{2}{\Omega_p}\left(\frac{P'}{\Sigma_p}+\Phi_p\right) + 3 x \bar{v}_y + \mathcal{O}\left(\frac{q^2}{h_\gamma^6}\frac{c_\gamma^2}{\Omega_p}\right).
\end{equation}
By evaluating the stream function at a stagnation point (located on the line $x = 0$ in the limit $q/h_\gamma^3 \to 0$), and then far up- or downstream on the same streamline, we see that
\begin{equation}
\frac{3}{4}\Omega_p x_s^2 = -\frac{2}{\Omega_p}\frac{q}{h_\gamma^3}c_\gamma^2 \chi_s,
\end{equation}
where $\chi_s$ is the value of the (non-dimensional) pseudo-enthalpy $\chi$ at a stagnation point on the separatrix streamline. From our numerical solution, we have
\begin{equation}
    \chi_s = -0.47115,
\end{equation}
and in the same limit $q/h_\gamma^3 \to 0$, the flow has three stagnation points, with coordinates $x = 0$, $y = \pm 0.439 H_\gamma$ and $y = 0$. We therefore predict a horseshoe region half-width for a low-mass planet of
\begin{equation}
    x_s = 1.12089\sqrt{\frac{q}{h_\gamma^3}}H_\gamma = 1.12089\sqrt{\frac{q}{h^3}}H\gamma^{-1/4},
\end{equation}
in good agreement with 2D simulations using a smoothing length $b = 0.4H_\gamma$ \citep[equation (44)]{paardekooper_torque_2010}, which match well the horseshoe width measured in 3D simulations for this choice of $b$ \citep{lega_outwards_2015,masset_horseshoe_2016}.

The exact $\gamma^{-1/4}$ dependence is a non-trivial result, arising from the coincidence that the 2D mode of the potential, specified in equation (\ref{Phidefn}), depends only on the length-scale $H_\gamma$ and not $H$, even though the vertical structure of the background disc varies vertically on the length-scale $H$. 

The phenomenon of three stagnation points appearing on the axis $x = 0$ seems to be physical, indeed it is observed in 3D simulations (see for example figures 3 and 4 in \citet{masset_horseshoe_2016}, where in their isothermal fiducial run, the stagnation points have approximate positions $y = - 0.36 H$, $y = 0.53 H$ and $y = 0$, having been displaced by the far-field radial pressure gradient).

In a 3D disc, in general the horseshoe width will depend on height. In the isothermal case however (when $\gamma = 1$), there is no entropy stratification, and in fact close to the orbital radius of the planet, the flow is columnar (akin to Taylor-Proudman columns) since the advection is dominated by Coriolis and body forces. In this way, the horseshoe width becomes approximately height-independent, as is observed in 3D isothermal simulations, for example those performed by \citet{fung_3d_2015} and \citet{masset_horseshoe_2016}.

In the adiabatic case, the Taylor-Proudman theorem no longer applies. Indeed, we expect a buoyancy wake within the downstream coorbital region, as observed by \citet{mcnally_low-mass_2020}. Instead, to gain traction here we appeal to the vertical structure of the 2D mode. As discussed in section \ref{s:td}, this mode behaves as
\begin{multline}
    v'_x \propto \exp{\left(\tfrac{(\gamma-1)z^2}{2 H_\gamma^2}\right)}, \quad v'_y \propto \exp{\left(\tfrac{(\gamma-1)z^2}{2 H_\gamma^2}\right)}, \quad v'_z = 0,\\
    p' \propto \exp{\left(-\tfrac{z^2}{2 H_\gamma^2}\right)}, \quad \rho' \propto \exp{\left(-\tfrac{z^2}{2 H_\gamma^2}\right)}.
\end{multline}
Consequently (if we neglect the contributions from the other vertical modes), we expect the horseshoe width to increase with height above the mid-plane as
\begin{equation}\label{HSz}
x_s(z) \approx 1.12 \sqrt{\frac{q}{h^3}}H \gamma^{-1/4}\sqrt{2-\gamma} \exp{\left(\tfrac{(\gamma-1)z^2}{2 H_\gamma^2}\right)}.
\end{equation}
The neglect of further 3D modes in this approximation is well-motivated. For example, inertial waves behave very differently to the 2D mode, with enthalpy perturbations typically modulated by $\sqrt{x}$ near to corotation \citep{latter_inertial_2009}. Gravity waves are also confined near to corotation to the region above diagonal buoyancy resonances \citep{lubow_analytic_2014}. 

Whilst the exponential increase in horseshoe width (as well as perturbed flow velocities) with height is striking, we note that that the factor $(\gamma - 1)/2\gamma$ is small for reasonable values of $\gamma$. Further, we shouldn't be too concerned with the dynamics beyond 2 to 3 scale heights above the mid-plane, as over 95\% of the disc's mass is contained within the first 2 scale heights. Moreover, our model's assumptions, including linearity, and that the background state is vertically isothermal (discussed in section \ref{s:caveats}) may begin to break down at such heights.

\begin{figure}\centering
  \includegraphics[width=0.99\linewidth]{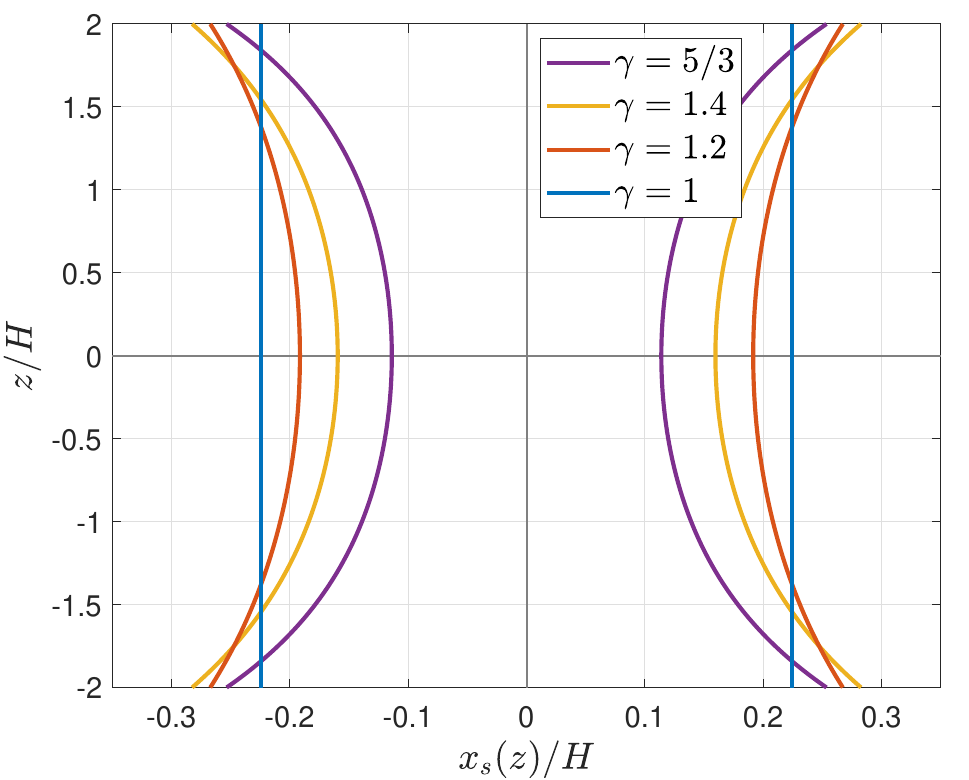}
\vspace{-1em}
\caption{Predicted vertical dependence of horseshoe region width $x_s(z)$ from 2D mode calculation for a planet with $M_p/M_\star = q = 5\times 10^{-6}$ embedded within a disc with aspect ratio $h = 0.05$.}
  \label{HSdep}
\end{figure}

Figure \ref{HSdep} depicts this vertical dependence for a few values of $\gamma$. Curiously, the expression in (\ref{HSz}) is poorly defined for $\gamma \geqslant 2$; however we don't expect $\gamma$ to take such values in astrophysical discs. This oddity is due to the change in the character of the the disc's linear modes as $\gamma$ increases through 2. For $\gamma \geqslant 2$, for example there is no mode with $v'_z = 0$ (such a mode would contain infinite energy). 

In the isothermal case, our numerical value for the horseshoe width $x_s = 1.12\sqrt{\frac{q}{h^3}}H$ is in reasonable agreement with 3D simulations (though inflated by a few percent). \citet{lega_outwards_2015} and \citet{masset_horseshoe_2016} both estimate a width via 3D simulation of $x_s = 1.05\sqrt{\frac{q}{h^3}}H$. The reasons for the discrepancy between our result and theirs are perhaps two-fold. Most significantly, we use a fully local expression for the potential (\ref{philoc}), rather than performing a discrete azimuthal mode decomposition that takes into account the finite circumference of the disc. In this way, our calculation is most applicable to ultra-thin discs (for example in the problem of black hole migration within an AGN disc). We expect a relative discrepancy between our value and those obtained in 3D simulations of order $h_\gamma \sim 0.05$. Furthermore, we exclude any non-linearity in our calculation. From equation (\ref{LOsf}) we expect a next order correction to $x_s$ of relative size $\mathcal{O}\left(q/h^3\right)$, which may also be on the order of a few percent. 

\subsection{Velocity profiles at corotation}

\begin{figure}
\centering
\includegraphics[width=0.99\linewidth]{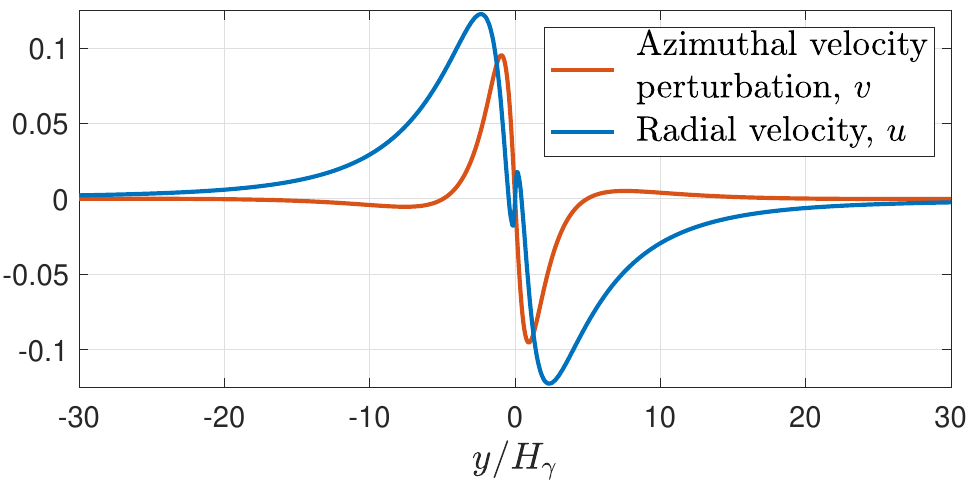}
\vspace{-1em}
\caption{Profiles of non-dimensionalised azimuthal and radial velocity perturbations at corotation, extracted from the numerical solution of (\ref{eqJpm0F}) and (\ref{eqvy0F}).}
  \label{crtvel}
\end{figure}

Of particular importance, especially when studying the dynamics of PV and entropy within the horseshoe region, is the dominant flow structure of the horseshoe streamlines. The radial extent of the horseshoe region is $\mathcal{O}\left(\sqrt{\frac{q}{h^3}}H\right)$, whereas the flow velocity varies on the length-scale $H$. As a result, the flow perturbations induced by the planet within the horseshoe region (which inform the dominant flow structure) are well approximated simply by their values on the line $x = 0$, that is, taking
\begin{equation}
    v_{y,\text{hs}} \approx -\frac{3}{2}\Omega_px + v'_y(0,y), \quad v_{x,\text{hs}} \approx v'_x(0,y)
\end{equation}
This was noted by \citet{balmforth_non-linear_2001}, who studied the saturation of the corotation torque, and used matched asymptotics to compute Fourier coefficients for the flow velocity perturbations near to corotation. Figure \ref{crtvel} depicts the non-dimensional radial and azimuthal velocity distributions at corotation, which satisfy (\ref{eqJpm0F}) and (\ref{eqvy0F}). These correspond to taking cross-sections of the flow depicted in figure \ref{3Disofigs1} along the line $x=0$.

\subsection{Wave evolution, profiles and one-sided torque}\label{s:waveevo}

\begin{figure*}\centering
  \includegraphics[width=0.495\linewidth]{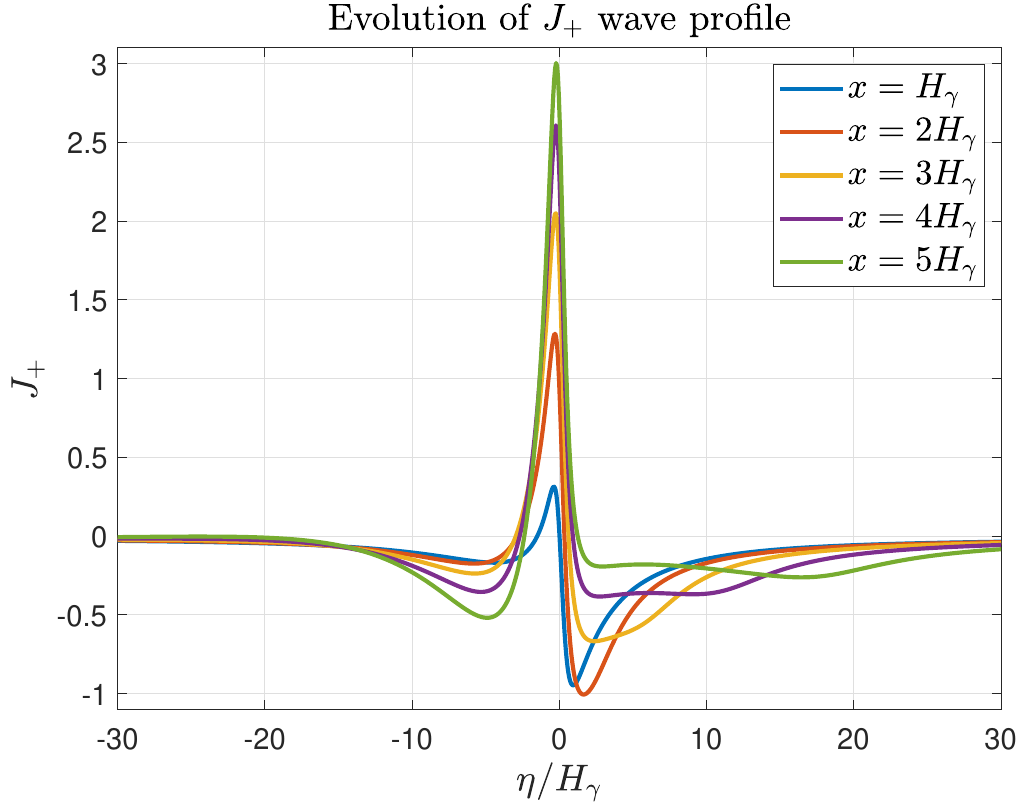}
  \includegraphics[width=0.495\linewidth]{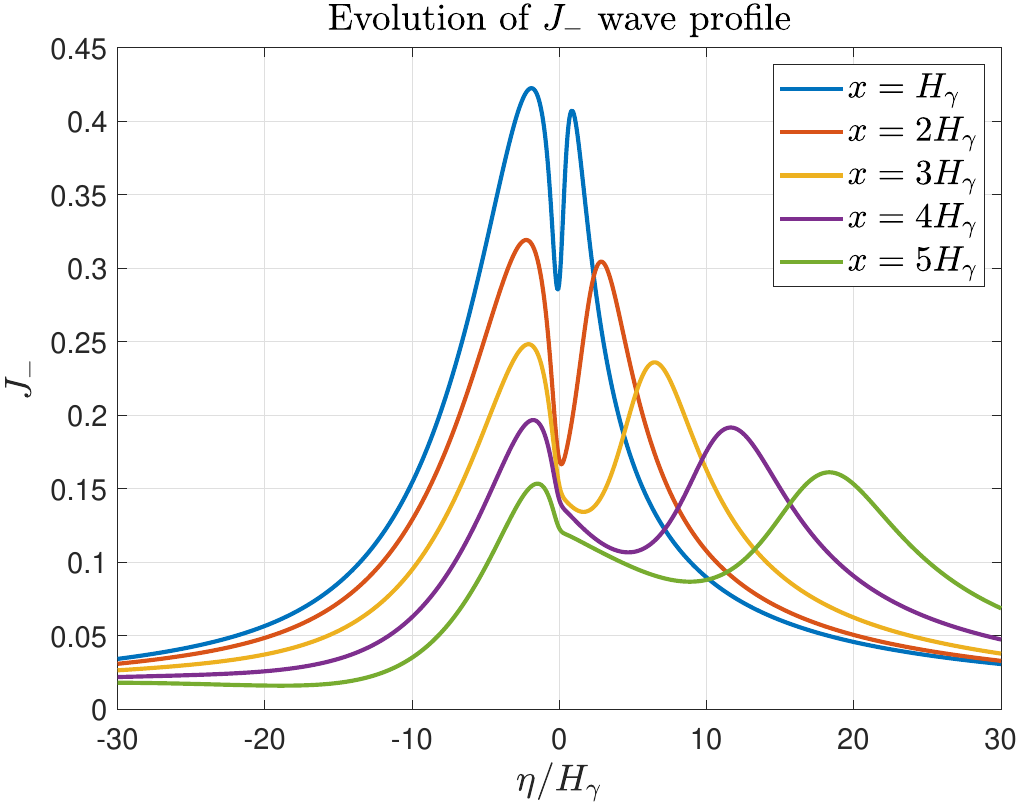}
  \includegraphics[width=0.495\linewidth]{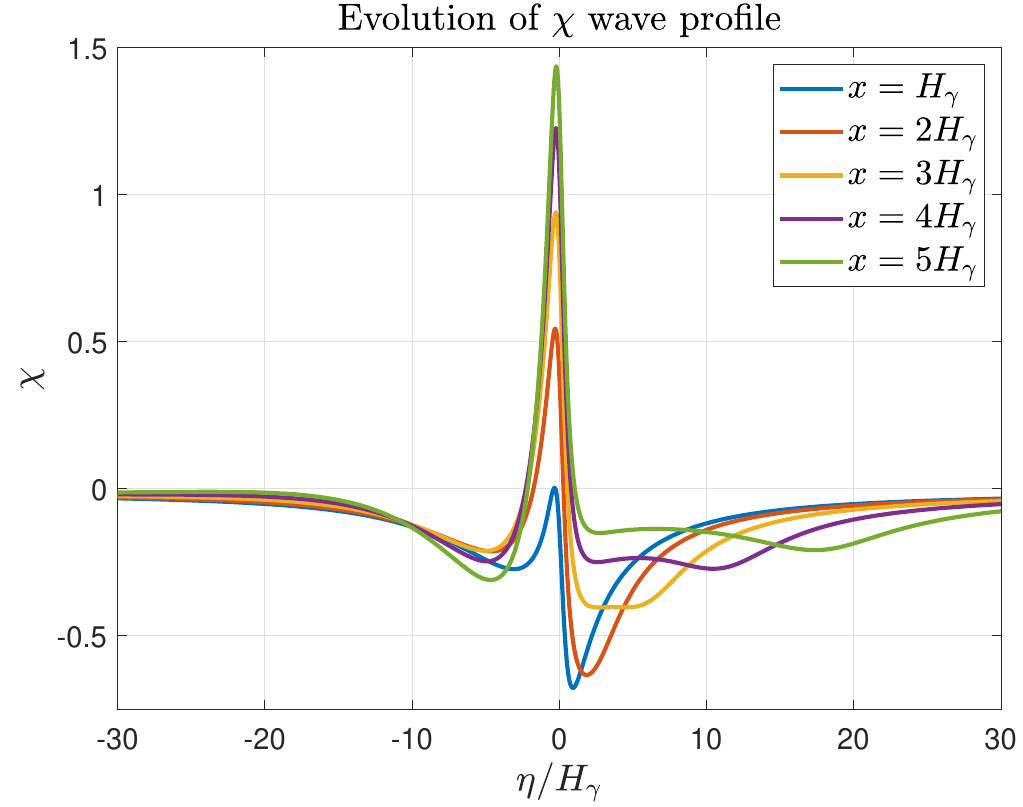}
  \includegraphics[width=0.495\linewidth]{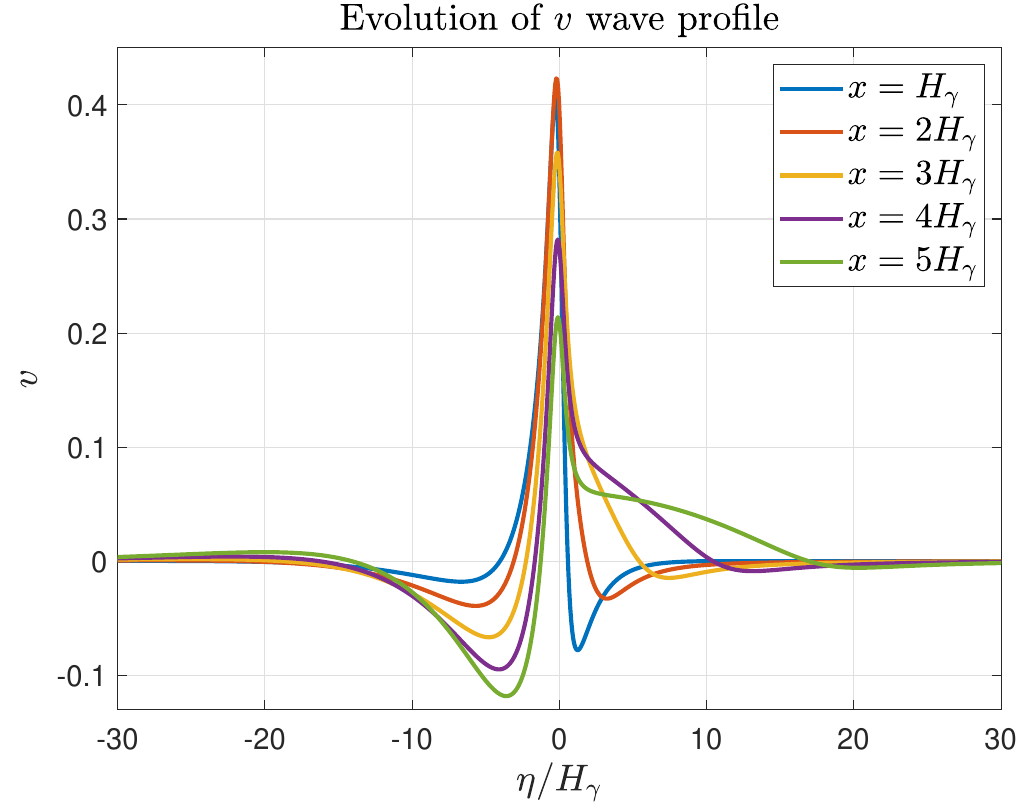}
  \caption{Profiles of non-dimensional linearized Riemann invariants $J_+$ and $J_-$, pseudo-enthalpy $\chi = W + \hat{\phi}_p$ and azimuthal velocity perturbation $v$ for $x/H_\gamma = 1$, $2$, $3$, $4$ and $5$. $\eta$ is the characteristic coordinate defined in equation (\ref{etadefn}). The scalings for the variables are given in equation (\ref{nondim}).}
  \label{profiles}
\end{figure*}

In this section we clarify in what sense $J_+$ and $J_-$ are linearized Riemann invariants, how they correspond to their approximately conserved non-linear counterparts, and discuss the excited wave profiles shown in figure \ref{profiles} and corresponding one-sided Lindblad torque.

The non-linear 2D theory developed by \citet{goodman_planetary_2001} neglects the azimuthal velocity perturbation, as its linear solution decays as $|x|^{-1/2}$ away from the planet. The resulting 2D system conserves on characteristic curves the Riemann invariants
\begin{equation}\label{RINL}
    R_\pm = u \pm \frac{2}{\gamma - 1}(c - c_\gamma), \quad c = c_\gamma\left(\frac{\Sigma}{\Sigma_p}\right)^{\frac{\gamma - 1}{2}}.
\end{equation}
For small departures from the background state, the enthalpy perturbation obeys $W'/c_\gamma \approx c - c_\gamma$. We may write therefore
\begin{equation}
    R_\pm \approx u \pm \frac{W'}{c_\gamma},
\end{equation}
equal to $J_\pm$ up to an unimportant additive contribution of $\Phi_p/c_\gamma$. It therefore makes sense to interpret $J_\pm$ as linearized Riemann invariants.

A WKB analysis, appropriately imposing outgoing wave boundary conditions, indicates that for each azimuthal Fourier mode (with wavenumber $k_y > 0$),
\begin{equation}
\tilde{J}_{+} \sim \begin{cases}
|x|^{+\tfrac{1}{2}}\exp{\left(+\ii\tfrac{3}{4}\tfrac{\Omega_p k_y}{c_\gamma}x^2\right)}, \quad \text{if}\;  x > 0\\
|x|^{-\tfrac{3}{2}}\exp{\left(-\ii\tfrac{3}{4}\tfrac{\Omega_p k_y}{c_\gamma}x^2\right)}, \quad \text{if}\;  x < 0.
\end{cases}
\end{equation}
The behaviour of $\tilde{J}_{-}$ can be determined from the symmetry $J_+(x,y) = -J_-(-x,-y)$, or in Fourier-space, $\tilde{J}_-(x,k_y) = -\tilde{J}_+^*(-x,k_y)$. Note the strong $|x|^{-3/2}$ decay of $\tilde{J}_{+}$ in $x < 0$. This may be thought of as a consequence of $J_+$ being approximately conserved on characteristics emanating from $x = -\infty$, where it is $0$ (so that we have a simple wave). The consequence of this can be seen in figure \ref{3Disofigs1}, where we see each Riemann invariant nearly vanish one side of the line $x = 0$. The decay is not quite as strong as $|x|^{-3/2}$ in real space close to the planet because of the influence of wave excitation there.

The spiral arm locations are well approximated by the characteristics of the second order operators in equations (\ref{eqJpm0F}) and (\ref{eqvy0F}). That is, the density waves in $x > 0$ follow the curve $\eta(x,y) = 0$, for characteristic coordinate
\begin{equation}\label{etadefn}
    \eta = y - \frac{x}{3}\sqrt{\frac{9}{4}x^2 - 1} - \frac{1}{3}\cosh^{-1}\left(\frac{3x}{2}\right).
\end{equation}
The accurate computation of these wave profiles is an important step in the determination of the total torque on the planet, contributing to the Lindblad torque. They're also needed to compute the locations of eventual shocking due to wave steepening. Whilst the profiles obtained within a strictly local approximation yield cancelling inner and outer torques, the modification to the profiles from any asymmetry will lead to a net torque on the planet. The profiles excited by the vertically averaged $\Phi_p$, shown in figure \ref{profiles}, are qualitatively very similar to the profiles obtained in planet-driven wave evolution studies \citep{goodman_planetary_2001,dong_density_2011}.

\begin{figure}\centering
  \includegraphics[width=0.99\linewidth]{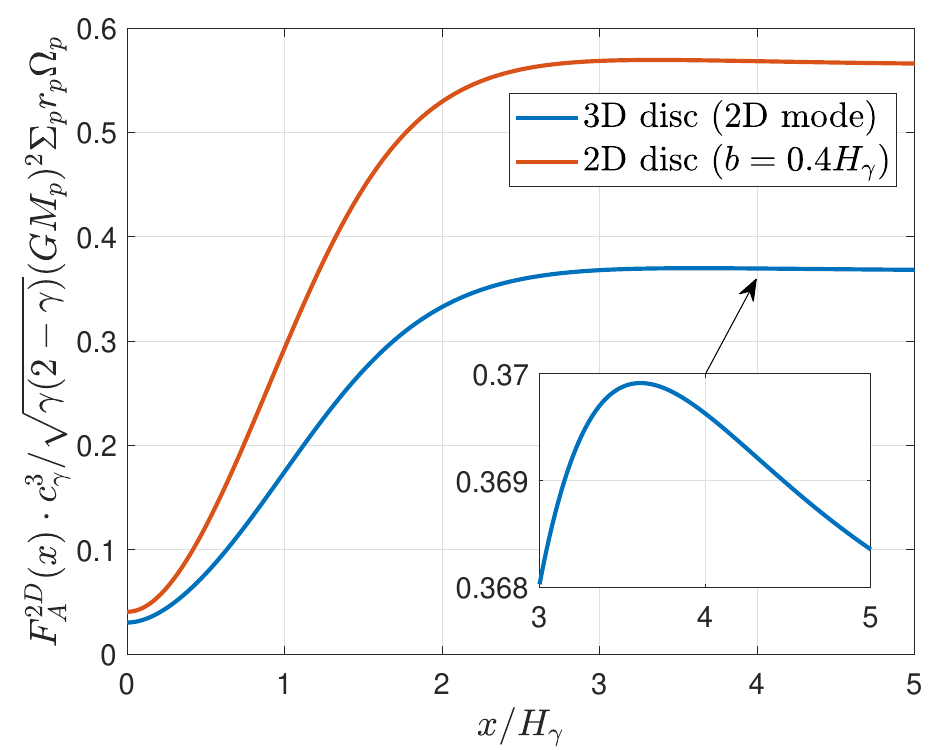}
\vspace{-1em}
\caption{The angular momentum flux, $F_A^{\text{2D}}(x)$, carried by the density wave in a 3D disc (calculated by projection onto the 2D mode, blue), compared with the flux excited in a 2D disc by a softened potential with smoothing length $b = 0.4H_\gamma$ (red). The box shows a zoomed in view of the flux profile, which exhibits the `negative torque phenomenon' \citep{rafikov_origin_2012}. The factor of $\sqrt{\gamma(2-\gamma)}$ in the $y$-axis scale originates from equation (\ref{fluxdecomp}), and may be ignored when interpreting the flux in the softened case.}
  \label{flux}
\end{figure}

Our one-sided torque estimate is however more than a factor of 2 smaller than that obtained in the 2D studies conducted by \citet{goodman_planetary_2001} and \citet{dong_density_2011}. We attribute this to the point-mass and softened potentials used to force the 2D system in previous studies artificially inflating the excited wave amplitudes. This is the case even for `higher order' expressions for the planet potential such as those used by \citet{dong_density_2011}, and especially for smaller softening parameters. Indeed, we find that the angular momentum flux carried by the 2D mode, $F_A^{\text{2D}} = \sqrt{\gamma(2-\gamma)}r_p\Sigma_p\int_{-\infty}^{\infty}\bar{v}_x\bar{v}_y \dd y$ (as defined in section \ref{s:td}), far from the planet is
\begin{equation}
F_A^{\text{2D}}(\infty) = 0.37 \sqrt{\gamma(2-\gamma)}\frac{(G M_p)^2 \Sigma_p r_p \Omega_p}{c_\gamma^3},
\end{equation}
corresponding to a one-sided torque
\begin{equation}\label{torque}
    T^{\text{2D}} \equiv F_A^{\text{2D}}\bigg|_0^{\infty} = 0.34 \sqrt{\gamma(2-\gamma)}\frac{(G M_p)^2 \Sigma_p r_p \Omega_p}{c_\gamma^3}.
\end{equation}
For comparison, in the case of a point-mass potential in a 2D disc, \citet{goldreich_disk-satellite_1980} calculate a dimensionless numerical torque prefactor of $0.93$. We've used here that $F_A^{\text{2D}}(0) = 0.03 \sqrt{\gamma(2-\gamma)}\frac{(G M_p)^2 \Sigma_p r_p \Omega_p}{c_\gamma^3}$, which is apparent from figure \ref{flux}. The fact that $F_A^{\text{2D}}(0) \neq 0$ is due to a small fraction (4\%) of the flux carried by the outward-propagating wave being excited in $x < 0$.

In the case $\gamma = 1$, the torque estimate (\ref{torque}) is in agreement with the 3D (isothermal) simulations of \citet{dangelo_three-dimensional_2010} to within a few percent (one can find a one-sided Lindblad torque estimate from the symmetric part of the cumulative torque profiles in their figures 7 and 8); it further matches precisely the torque found by \citet{zhu_planet-disk_2012}. This is because in the case $\gamma = 1$, all gravity waves are suppressed, and inertial waves carry very little flux \citep{tanaka_threedimensional_2002}.

\citet{zhu_planet-disk_2012} and \citet{arzamasskiy_three-dimensional_2018} also note the discrepancy of a factor of 2 or 3 between previous 2D and 3D estimates of the one-sided torque. Judging from our result and figure 2 of \citet{zhu_planet-disk_2012}, it appears that in the case $\gamma = 1.4$, the 2D mode contributes only 75\% of the one-sided Lindblad torque, with inertial waves and gravity waves carrying the majority of the remaining flux.

The angular momentum flux computed from our numerical solution, $F_A^{\text{2D}}(x)$, is plotted in figure \ref{flux}. Of note is that the torque is almost entirely exerted within a few scale heights of the planet, which justifies a local approach to the problem. Further, there's a very small decrease in the flux beyond $x \approx 3.5 H_\gamma$, which may be attributed to the `negative torque phenomenon' \citep{rafikov_origin_2012}.

Finally, we're able to comment on the likely vertical profile of the shock front, and offer insight into the modification to the shocking length incorporating 3D effects. We note that for a freely propagating 2D mode, $v'_z = 0$, so that wave steepening may be considered as a purely horizontal effect. Further, the density wave wake and the gravity wave wake are spatially separated far from the planet \citep{mcnally_low-mass_2020}, which suggests considering the wave steepening of an isolated 2D mode would provide a good model for 3D shock formation.

Using equation (\ref{2dzprof}) and comparing the maximum steepness of our 2D mode profile with that of \citet{goodman_planetary_2001} and following their wave steepening calculation, we estimate the location of the initial shock in a 3D disc relative to their 2D estimate as
\begin{equation}
    l_\text{sh}^{\text{3D}} \approx 1.2\frac{\ee^{-\tfrac{\gamma - 1}{5 H_\gamma^2}z^2}}{(2-\gamma)^{1/5}}\cdot l_\text{sh}^{\text{2D}}.
\end{equation}
Here, we've simply exploited the $A^{-2/5}$ dependence of the shocking length on the amplitude of the wave, and noted the reduced maximum steepness of our linear 2D mode surface density profile (which is around 65\% of that found by \citet{goodman_planetary_2001} at $x = 1.33 H_\gamma$). The vertical dependence is weak for physical values of $\gamma$; for $\gamma = 1.4$ it varies by only 5\% within the first scale height, $H_p$, above the disc mid-plane, though higher up a shock may be expected to form slightly closer to the planet. At much greater height however, as discussed in section \ref{s:flowsl}, our model's basic assumptions may begin to break down.

\section{Discussion}
\label{s:discussion}

We've already discussed many key items as and when they arose earlier in this paper, including
\begin{itemize}
    \item the nature and resolution of the corotation singularity in section \ref{s:crtsing}, including the manner in which it has a `non-linear' resolution, and how and why it appears in previous linear analyses;
    \item the relation between the commonly adopted 2D Euler equations and variables and their 3D counterparts (section \ref{s:2D});
    \item the appearance of and displacements in the locations of the Lindblad resonances within the framework of this paper in section \ref{s:LR}.
\end{itemize}
However, it's worth discussing caveats to this work as well as two further topics which warrant additional attention: the `rigorous' softening of planetary potentials to reproduce 3D effects within a 2D disc model, and the role that this 2D mode plays in determining the torque on a low-mass planet. 

\subsection{Treating planetary potentials in 2D discs}\label{s:softening}

\subsubsection{Context: softening prescriptions}

When studying planet-disc interactions in 2D, the planet's potential should be modified to account for the influence of the vertical extent of the disc on the dynamics. Prescriptions such as
\begin{equation}\label{bsoft}
    \Phi_p = -\frac{G M_p}{\sqrt{\left|{\bf{r}} - {\bf{r}}_p\right|^2 + b^2}}, \quad b \sim H/2,
\end{equation}
are commonly adopted, as well as other similar prescriptions including the `fourth order' softened potential described in equation (16) of \citet{dong_density_2011}. Each of these prescriptions however struggle to uniformly capture the impact of the disc's vertical extent on the dynamics within a few scale heights of the planet.

\citet{muller_treating_2012} compared (\ref{bsoft}) to the density-weighted average of a point-mass potential and concluded that the best choice for $b$ varies depending on the distance from the planet, especially for separations $\left|{\bf{r}} - {\bf{r}}_p\right| \sim H$. They note, worryingly, that the choice of $b$ can inform the direction of planetary migration, or in fragmentation simulations whether the disc fragments or not. For our purposes, adopting such a prescription can significantly inflate the excited wave flux (by as much as or more than a factor of 2, for both (\ref{bsoft}) and the `fourth order' potential), and can impact strongly the horseshoe region width. As mentioned previously, the variation of net torque components can be by more than a factor of 3 as smoothing length is varied between $0.3$ and $0.7$ (\citet{tanaka_three-dimensional_2024}, figure 5).

\subsubsection{A rigorous choice}

In light of this, in this section we emphasise that (\ref{Phidefn2}) is a far better choice for the 2D potential. In particular, taking inspiration from \citet{lubow_wave_1993}, we proved in section \ref{s:proj} that in the case of a low-mass planet, it is possible to directly derive 2D fluid equations governing averaged velocities and an effective `surface density' which are forced precisely by (\ref{Phidefn2}). In this way, we argue that (\ref{Phidefn2}) (which for simplicity excludes the height-independent indirect term) is the optimal choice for the planetary potential in a 2D disc, especially when considering sub-thermal mass planets.
\begin{equation}\label{Phidefn2}
    \Phi_p \equiv -\frac{G M_p}{H_\gamma}\frac{\ee^{\frac{1}{4}s^2}}{\sqrt{2\upi}}K_0\left(\tfrac{1}{4}s^2\right), \quad s = \left|{\bf{r}}-{\bf{r}}_p\right|/H_\gamma.
\end{equation}
(We discuss below how this may be implemented practically in simulations.)

Indeed, (\ref{Phidefn2}) is uniformly a very good approximation to the potential which forces the \emph{global} 2D mode of the disc (which would simply involve $s = \left|{\bf{r}}-{\bf{r}}_p\right|/H_\gamma(r)$), since far away from the planet
\begin{equation}
    \Phi_p \sim -\frac{G M_p}{\left|{\bf{r}}-{\bf{r}}_p\right|}\left[1 - \frac{1}{2s^2} + \frac{9}{8s^4} + \mathcal{O}\left(s^{-6}\right)\right].
\end{equation}
That is, far away the potential returns to the familiar point-mass potential with small correction terms. In the same way, we reassuringly recover the point mass potential in the case of a `razor-thin' disc. The higher order corrections may be thought to arise physically from the quadrupolar and higher order multipolar interactions between the vertically extended disc and the planet monopole. In seeking to converge more rapidly to a point-mass potential, the `fourth order' softened potential of \citet{dong_density_2011} neglects the quadrupolar interaction term.

Near the planet, if we assume the background temperature $T_0(r)$ (and therefore the scale height $H(r)$) varies slowly with radius, then the departure of the potential which forces the global 2D mode from the above expression for $\Phi_p$ is also small.

\begin{figure*}\centering
  \includegraphics[width=0.495\linewidth]{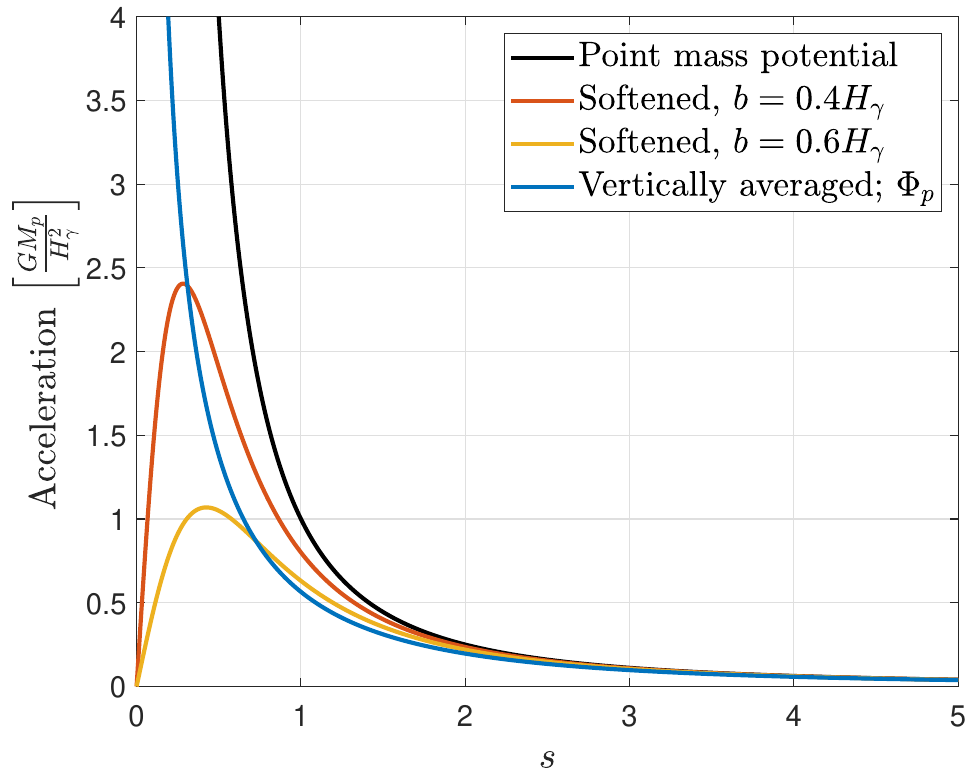}
  \includegraphics[width=0.495\linewidth]{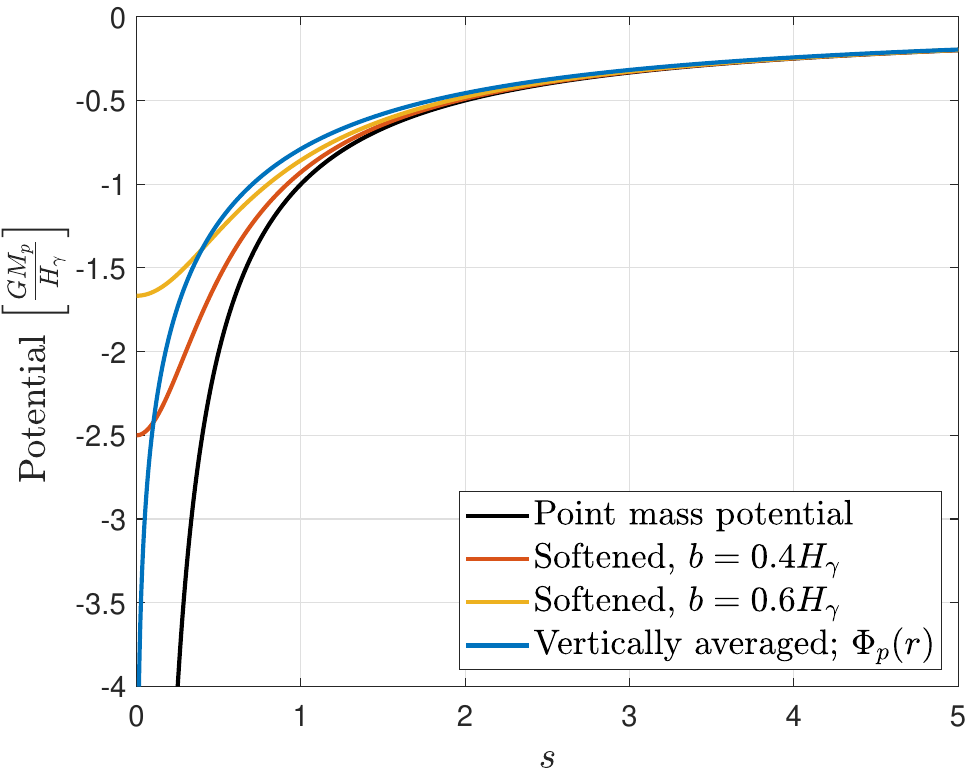}
  \caption{Comparison of vertically averaged potential $\Phi_p$ from equation (\ref{Phidefn}) which forces the 2D equations (\ref{2Dxmom}) and (\ref{2Dymom}) with point-mass and softened potentials of the form (\ref{bsoft}). The left panel shows acceleration for each prescription and the right panel shows gravitational potential. $s = \left|{\bf{r}}-{\bf{r}}_p\right|/H_\gamma$ measures the in-plane distance from the planet in adiabatic scale heights.}
  \label{Phicomp}
\end{figure*}
In figure \ref{Phicomp} we compare graphically the appearance of the vertically averaged potential (\ref{Phidefn2}) with a point-mass potential and `softened' potentials of the form (\ref{bsoft}), as well as their corresponding gradients. Note that $\Phi_p$ has a logarithmic singularity at the origin. More precisely,
\begin{equation}
    \Phi_p \sim \frac{G M_p}{H_\gamma} \frac{2 \ln s}{\sqrt{2\upi}} \quad \text{as} \; s \to 0.
\end{equation}
A logarithmic singularity is a general feature of any weighted vertical average of the planetary potential.

Despite the singularity experienced by the averaged potential (and correspondingly the enthalpy) at the location of the planet, the linearity of the system is hardly compromised. To demonstrate this, we note that in reality, the singularity of the 3D potential will be truncated at the planet's radius, $R_p$. We define the small parameter
\begin{equation}
    \varepsilon = R_p/H_\gamma.
\end{equation}
For Earth-like planets located at around $1$ to $10$ au, we expect $10^{-5} \lesssim \varepsilon \lesssim 10^{-3}$. Suppose we truncate the singularity of the 3D planetary potential (excluding the indirect term) at the planet's radius prior to the vertical averaging as
\begin{equation}
    \phi_p = \begin{cases}
        -\frac{G M_p}{\left|{\bf{r}}-{\bf{r}}_p\right|}, \quad \left|{\bf{r}}-{\bf{r}}_p\right| > R_p,\\
        -\frac{G M_p}{R_p}, \quad \left|{\bf{r}}-{\bf{r}}_p\right| \leqslant R_p.
    \end{cases}
\end{equation}
In this case,
\begin{equation}
    \Phi_p({\bf{0}}) \sim \frac{G M_p}{H_\gamma}\frac{2\ln\varepsilon}{\sqrt{2\upi}}.
\end{equation}
However, $\left|2\ln\varepsilon/\sqrt{2\upi}\right| \sim 7 < h_\gamma^3/q$ by assumption, so that the 2D system (including the enthalpy $W$) everywhere constitutes an approximately linear perturbation to the background state. We note further that the solutions for $u$, $v$ and $\chi$ shown in section \ref{s:results} are very weakly affected by the (integrable) logarithmic singularity in the potential.

This then suggests that for numerical applications it would be very reasonable (namely to avoid numerical divergences and timestep issues), to `soften' the 2D potential (\ref{Phidefn2}) by taking $s^2 = \varepsilon^2 + \left|{\bf{r}}-{\bf{r}}_p\right|^2/H_\gamma^2$, though here a practical choice for $\varepsilon$ for grid-based simulations in which the planetary radius is unresolved would be on the order of the grid spacing divided by $H_\gamma$.

Regarding numerical implementation, we ought to address the subtlety that when directly evaluating the potential for large arguments $s$, we multiply both exponentially large and small numbers. However, efficient C functions for the exponentially scaled Bessel functions $\ee^xK_0(x)$ and $\ee^xK_1(x)$ (which both appear in the expression for the gradient of $\Phi_p$) already exist within the Cephes library\footnote{\url{https://www.netlib.org/cephes/}} \citep{moshier_methods_1989}. Alternatively, the vertically averaged potential in (\ref{Phidefn2}) and its derivative are functions of one parameter only, so that a lookup table would provide an efficient method for evaluation.

\subsection{Torques on low-mass planets}

The torque exerted on the planet by the disc is of particular importance, as it informs the rate and direction of planet migration. The solution presented in this paper constitutes the leading order flow induced by the planet, which is rotationally symmetric and so exerts no net torque on the planet. The next-order correction to this flow is a factor of $h_\gamma$ smaller, includes geometric asymmetries as well as radial variations in the properties of the background disc, and in general leads to a net torque on the planet. The majority of the torque exerted on the planet is done so locally. In this way, a radially local model of the planet-disc interaction is sufficient to determine the torque on the planet, which must depend linearly on the local gradients of the disc's properties.

\subsubsection{Corotation torque and critical layer at corotation}

One key complicating issue is that of the horseshoe streamlines. They form closed loops when the full azimuthal extent of the flow is considered (and when the planet and disc are not migrating or drifting radially relative to each other). In particular, materially conserved quantities (such as Ertel's PV and the entropy) are advected from inner to outer disc radii, and vice versa. In the absence of strong viscosity or dissipation therefore, the properties of the background disc are not restored far (azimuthally) from the planet.

This effect gives rise to an additional torque known as the corotation torque. To determine this torque, it's necessary first to solve for the distributions of these conserved quantities within the critical layer which is called the horseshoe region, with appropriate prescriptions for viscosity, heating etc. A key ingredient in this solution is an accurate description of the leading order flow within the corotation region. As mentioned in the \hyperref[s:intro]{introduction}, there has already been a wealth of research devoted to the study of corotation torques. However, there is still more to be explored here, including in particular the impact that the vertical extent of the disc has on the torque. Part of this impact is captured in the behaviour of the 2D mode of the leading order flow, whose velocity profiles at corotation are shown in figure \ref{crtvel}. These profiles vary only slightly over the corotation region, and may be used to enable cheap numerical studies of the influence of a range of physics on the corotation torque.

\subsection{Caveats}\label{s:caveats}

We point out three main caveats to this work, which are detailed in the subsections below.

\subsubsection{Omitted 3D wave modes}

The solution presented in this paper is not fully 3D. We solved for the behaviour of the $z$-dependent 2D mode of the disc, which ought to be superposed with a spectrum of inertial and internal gravity waves. These contribute a significant fraction (upwards of 20\% for $\gamma = 1.4$) of the one-sided torque on the planet, as discussed in section \ref{s:waveevo}.

\subsubsection{Impact of turbulence and magnetic fields}

Many physical phenomena may act to disrupt the solution for the disc's 2D mode. The vertical shear in the 2D mode might be affected by processes that provide greater coupling between different layers in the disc, for example turbulence, viscosity, or a vertical magnetic field. Favourable conditions include a large plasma $\beta \gg 1$ and low level of ionisation, as well as a low Reynolds stress, captured by the condition for the viscosity parameter $\alpha \ll 1$ \citep{shakura_black_1973}. It's widely believed that except for the innermost regions, protoplanetary discs do predominantly exhibit $\beta \gg 1$ \citep{lesur_2021}, and recent numerical and observational studies point towards a smaller turbulent viscosity than previously thought. Non-ideal MHD effects such as ambipolar diffusion and Ohmic resistivity act to suppress turbulence even in the disc's surface layers \citep{turner_2014}, and to decouple the fluid from the magnetic field. Indeed, \citet{pinte_dust_2016} and later \citet{villenave_highly_2022} find that observations of the discs around HL Tau and Oph163131 are consistent with small viscosity parameters $\alpha \sim 10^{-4}$ and $\alpha \sim 10^{-5}$ respectively. 

\subsubsection{Thermodynamic assumptions}\label{s:disctherm}

The validity of the thermodynamic model adopted in this paper (which simply constitutes a vertically isothermal background state permitting adiabatic perturbations) depends critically on the thermal relaxation time-scale for the disc's gas. If the cooling is too rapid, then the adiabatic assumption breaks down, too slow and the disc's background state won't have reached a thermal equilibrium as the disc evolves. The disc's outer layers are typically hotter than its interior due to the stellar irradiation, which is not able to penetrate the optically thick disc interior. The molecular gas which comprises the disc interior is optically thick to its own radiative emission; its primary means of cooling is via the surrounding dust grains. The gas must first transfer its thermal energy to the dust, which is then able to radiate away the excess energy more efficiently (though the disc is not necessarily optically thin to this emission). As pointed out by \citet{malygin_efficiency_2017}, \citet{barranco_zombie_2018}, \citet{pfeil_mapping_2019} and \citet{bae_observational_2021}, infrequent gas-dust collisions often act as a bottleneck for the thermal relaxation of the gas, especially in its surface layers. \citet{bae_observational_2021} predict that typical gas thermal relaxation time-scales are tens of orbits even at 10 au, corresponding to a cooling parameter $\beta \equiv 2\upi t_{\text{relax}}/t_{\text{orb}} \gtrsim 10^{1.5}$.

This slow cooling offers favourable conditions for the adiabatic model governing planet-induced perturbations to represent a good approximation, though the cooling is still more rapid than the disc's evolution time-scale. Indeed, the downstream buoyancy wake excited by a giant planet orbiting at 90 au is believed to be the source of the tightly wound spirals observed in TW Hya \citep{teague_spiral_2019,bae_observational_2021}. The existence of such a signature adds to the credibility of the adiabatic model, as it implies $t_\text{relax} \gg N_z^{-1}$ (for $N_z$ the Brunt-Väisälä frequency), necessary for the buoyancy wake to develop. It's likely however that cooling and thermal diffusion play important roles on the much longer time-scale associated with the librating horseshoe motion. Whilst this has only a small effect on the dominant flow behaviour found in this paper, it has important consequences for the entropy and potential vorticity distributions within the horseshoe region, and correspondingly implications for the corotation torque.

Insight into the vertical temperature structure of the outer regions of protoplanetary discs is offered by the emission from different CO isotopes which trace different disc altitudes. Observational studies typically find a flat temperature plateau near the disc mid-plane, with an increase in temperature in the disc's upper layers, a few pressure scale heights above the mid-plane \citep{dartois_structure_2003,law_mapping_2024}. This is consistent with the physical picture discussed above. The disc's surface temperature is typically a factor of 2 larger than its mid-plane temperature. Whilst there's no exact analogue for the 2D projection procedure described in section \ref{s:proj} when the temperature of the background state varies with height, the majority of the disc's mass is contained within the region of temperature plateau. Whilst an oversimplification, this suggests therefore that the vertically isothermal model for the background disc may be expected to give quantitatively reasonable results.

\section{Conclusions}
\label{s:conclusions}

For linear adiabatic perturbations to a vertically isothermal protoplanetary disc, there is a particular vertical average of the 3D gas dynamic equations which exactly yields the familiar 2D linear system, comprised of averaged flow velocities, but an effective surface density, pressure, and planetary potential. This averaging process is closely related to the projection operator onto the Lubow-Pringle 2D mode of the disc \citep{lubow_wave_1993}. Importantly, when compared with more traditional softening prescriptions for planetary potentials, adopting the averaged potential (\ref{Phidefn}) provides a more rigorous and accurate, parameter-free method to modify the potential of an embedded planet to account for 3D effects.

In this paper we presented the solution for the 2D mode of the flow excited by a low-mass planet on a circular orbit, whose features include a spiral wake as well as horseshoe streamlines within the coorbital region. We derived non-singular, independent second order equations for each flow variable, (\ref{eqJpm0F}) and (\ref{eqvy0F}). These include novel parabolic cylinder equations for linear combinations of the radial velocity and enthalpy, and offer a correction to the model of the coorbital flow proposed by \citet{paardekooper_width_2009}.

The 2D mode (so-called as it has the property $v_z = 0$, though in general it is $z$-dependent) is a member of the wider family of 3D disc modes. It provides an interpretation for 2D disc models, and plays a dominant role in the response of the disc to the planet, particularly at corotation. We find that in the limit of an ultra-thin disc, the vertically averaged horseshoe width is $x_s = 1.12\sqrt{\frac{q}{h^3}}H\gamma^{-1/4}$. Taking only the 2D mode contribution to the flow predicts a horseshoe width which grows with height above the disc mid-plane as described in equation (\ref{HSz}) and figure \ref{HSdep}. 

The flow in the corotation region is well approximated by superposing the background shear flow with the perturbed flow fields evaluated at corotation, $x = 0$. This coorbital flow solution, depicted in figure \ref{crtvel}, may be used to inexpensively simulate the impact of various physics on the corotation torque, including diffusive and migration-feedback effects.

Our approach also allowed us to capture accurately the wave angular momentum flux transported by the spiral density wave in a 3D disc, which is reduced by a factor of 2 or 3 compared with previous 2D estimates. We used the profiles of this wave to estimate the location at which it first shocks in a 3D disc, as a function of height above the disc mid-plane.

We demonstrated that the 2D mode is orthogonal to the wider family of permitted 3D motions, which include for example gravity waves and inertial waves. As such, the torque on the planet decomposes into the sum of separate 2D and 3D components. We omit from this paper any discussion of the remaining excited spectrum of 3D wave modes. Resolving and understanding the gravity wave spectrum is a challenging and under-studied, yet important problem which requires future attention, not only because of the angular momentum which they transport, but also for their observational signature.

\section*{Acknowledgements}

This research was supported by an STFC PhD studentship (grant number 2750631). We are very grateful to the referee, Cl\'ement Baruteau, for a thorough report and very helpful suggestions which have improved the paper.

\section*{Data availability}

The data underlying this article will be shared on reasonable request to the corresponding author.

\bibliographystyle{mnras}
\bibliography{bibmaster}

\section*{Appendix A: numerical procedure}\label{appx}

\renewcommand{\theequation}{A\arabic{equation}}
\setcounter{equation}{0}

In this section we outline how the numerical solutions plotted in figure \ref{3Disofigs1} were obtained. We sought solutions with very high resolution and accuracy so that in future work we may use them to predict accurately the intricate dynamics which inform the corotation (and indeed wave) torques. We consider the problem in the $(x,k_y)$ plane, having Fourier transformed with respect to $y$. For brevity we denote $k_y = k$. It is helpful to define
\begin{equation}\label{pcedefs}
x' = \sqrt{3|k|}x, \quad a = \frac{1 + k^2}{3 |k|}.
\end{equation}
Equations (\ref{eqJpm0F}) and (\ref{eqvy0F}) become the canonical parabolic cylinder equations,
\begin{subequations}\label{0numeqs}
\begin{align}
\begin{split}
&\left[\upartial_{x'}^2 + \tfrac{1}{4}x'^2 - (a \pm \ii \; \text{sgn}(k))\right]\tilde{J}_{\pm} \\ &\qquad \qquad = - \text{sgn}(k)\left[\tfrac{1}{3}\ii + \tfrac{1}{2}\ii x'\upartial_{x'} \mp \left(\tfrac{1}{4}x'^2 - \tfrac{1}{3 k}\right)\right]\tilde{\phi}_p.
\end{split}
\\
&\left[\upartial_{x'}^2 + \tfrac{1}{4}x'^2 - a\right]\tilde{v} = \frac{1}{2\sqrt{3}}\left[x'\sqrt{|k|} - \tfrac{1}{\sqrt{|k|}}\upartial_{x'}\right]\tilde{\phi}_p,
\end{align}
\end{subequations}
forced by the Fourier transform of the potential $\hat{\phi}_p$, which we denote $\tilde{\phi}_p$. It may be shown (namely by evaluating the Fourier transform prior to the vertical averaging) that
\begin{equation}
\tilde{\phi}_p(x,k) = -\sqrt{\frac{2}{\upi}}\int_{-\infty}^{\infty}K_0\left(|k|\sqrt{x^2+z^2}\right)\ee^{-z^2/2}\dd z,
\end{equation}
which is far cheaper to evaluate accurately than the direct expression for the transform of $\hat{\phi}_p$.

Now, as $x' \to +\infty$, the solution will consist of the homogeneous waves $U\left(\ii a',x'\ee^{-\ii\frac{\upi}{4}}\right)$ and $U\left(-\ii a',x'\ee^{\ii\frac{\upi}{4}}\right)$. Here $a'$ is understood to denote the order of the relevant parabolic cylinder equation, that is, either $a\pm \ii$ or $a$, and we adopt $U(a,x)$ as defined in chapter 19 of \citet{abramowitz_handbook_1972}. We use the parabolic cylinder function (PCF) $U$, as $E(a',x)$ and $W(a',x)$ are not defined in general for $a' \in \mathbb{C}$. From now on we take $k \geqslant 0$, noting our {\emph{real}} real-space variables will have conjugate-symmetric Fourier transforms. With this in mind, and noting that $\forall \delta > 0$, the PCF $U(a,z) \sim z^{-a-1/2}\ee^{-\frac{1}{4}z^2}$ as $z \to \infty$ in $\arg(z) \leqslant \tfrac{3}{4}\upi - \delta$, we identify the following solutions as in- or outgoing waves:
\begin{subequations}
\begin{align}
&U\left(\ii a',x'\ee^{-\ii\frac{\upi}{4}}\right) & \text{is an outgoing wave  as} \; x' \to \infty\\
&U\left(-\ii a',x'\ee^{\ii\frac{\upi}{4}}\right) & \text{is an ingoing wave as} \; x' \to \infty\\
&U\left(\ii a',-x'\ee^{-\ii\frac{\upi}{4}}\right) & \text{is an ingoing wave as} \; x' \to -\infty\\
&U\left(-\ii a',-x'\ee^{\ii\frac{\upi}{4}}\right) & \text{is an outgoing wave as} \; x' \to -\infty
\end{align}
\end{subequations}
We want no incoming waves present in the solution. That is, our desired particular integral solution for $\tilde{J}_\pm$ or $\tilde{v}$, (which we denote $\tilde{\eta}_p$) behaves as
\begin{subequations}
\begin{align}
&\tilde{\eta}_p \sim b_1U\left(\ii a',x'\ee^{-\ii\frac{\upi}{4}}\right) \quad \text{as} \; x' \to \infty\\
&\tilde{\eta}_p \sim b_2U\left(-\ii a',-x'\ee^{\ii\frac{\upi}{4}}\right) \quad \text{as} \; x' \to -\infty.
\end{align}
\end{subequations}
We eliminate incoming waves in our numerical solution with the following algorithm. Suppose we compute a numerical solution $\tilde{\eta}$, imposing arbitrary initial data at $x' = 0$. It follows that, without loss of generality
\begin{subequations}
\begin{align}
\begin{split}
&\tilde{\eta} \sim (b_1 + c_1) U\left(\ii a',x'\ee^{-\ii\frac{\upi}{4}}\right) \\& \qquad\qquad\qquad + c_2 U\left(-\ii a',x'\ee^{\ii\frac{\upi}{4}}\right) \quad \text{as} \; x' \to \infty,
\end{split}
\\
\begin{split}
&\tilde{\eta} \sim (b_2 + d_1)U\left(-\ii a',-x'\ee^{\ii\frac{\upi}{4}}\right) \\&\qquad\qquad\qquad + d_2U\left(\ii a',-x'\ee^{-\ii\frac{\upi}{4}}\right) \quad \text{as} \; x' \to -\infty.
\end{split}
\end{align}
\end{subequations}
 We may then fit the numerical solutions via linear regression to the homogeneous PCF wave solutions far from $x' = 0$ (where the forcing has decayed sufficiently) to extract values for $c_2$ and $d_2$. We note that $c_2$ and $d_2$ both depend linearly on the (arbitrary) initial data imposed at $x' = 0$. Integrating the system twice more with two different choices for initial data allows us to solve this linear system for the correct initial data (which yields $c_2 = d_2 = 0$). We then perform one final integration with this choice of initial data, giving a numerical solution for $\tilde{\eta}_p$. 

 We carried out this algorithm with a 4th order Runge-Kutta ODE solver to compute the solutions to equations (\ref{0numeqs}). This method provides a very high numerical accuracy, and we were able to achieve a relative error of no more than $1\times 10^{-5}$ for all $0.01 \leqslant k \leqslant 500$.

It's worth noting some explicit analysis concerning the $k=0$ case, which is not permitted in the previous algorithm, and further how we can treat the singularities in the system in both real and Fourier-space without adopting softening prescriptions or cutoffs. In the case $k=0$, $\tilde{\phi}_p$ doesn't exist, and indeed $\tilde{J}_\pm$ diverges too. However, since
\begin{equation}
\left[1-\upartial_x^2\right]\tilde{W}\Big|_{k=0} = \upartial_x^2\tilde{\phi}_p, \;\; \text{and} \;\; \left[1-\upartial_x^2\right]\tilde{v}\Big|_{k = 0} = \frac{1}{2}\upartial_x\tilde{\phi}_p,
\end{equation}
for $\tilde{W}$ the transformed enthalpy perturbation, we may write
\begin{equation}\label{Wk0}
\tilde{W}(x,0) = \int_{-\infty}^{\infty}\ee^{-\left|x'-x\right|} \left[\sqrt{\frac{\upi}{2}}|x'|\text{erfcx}{\left(\frac{|x'|}{\sqrt{2}}\right)} - 1\right]\dd x',
\end{equation}
\begin{equation}
\tilde{v}(x,0) = \frac{1}{2} \int_{-\infty}^{\infty}\ee^{-\left|x'-x\right|} \sqrt{\frac{\upi}{2}}\text{erfcx}{\left(\frac{|x'|}{\sqrt{2}}\right)}\text{sgn}{(x')}\dd x',
\end{equation}
\begin{equation}
\tilde{u}(x,0) = 0,
\end{equation}
where $\text{erfcx}(x)$ is the scaled complementary error function. Now, as well as the divergence of the potential as $k \to 0$, we have that for large $k$,
\begin{equation}
\tilde{\phi}_p(x,k) \sim - \sqrt{2 \upi}\frac{\ee^{-|k x|}}{|k|},
\end{equation}
that is, as we should expect, the inverse Fourier transform of $\tilde{\phi}_p$ diverges at $x = y = 0$ (as the 2D real-space potential has a logarithmic singularity there).

We may however eliminate the need to evaluate the solutions at very many large values of $k$ (in order to obtain high accuracy), and overcome the singular behaviour at $k = 0$, by defining variables which are well-behaved everywhere in Fourier and real space. The inverse Fourier transform of these variables will then quickly give accurate results. First note that
\begin{equation}
\int_{-\infty}^{\infty}\ln(x^2+y^2)\ee^{-\ii k y}\dd y = -\frac{2 \upi}{|k|}\ee^{-|k x|}.
\end{equation}
It therefore makes sense to define regularised $J_\pm$ variables
\begin{equation}
J_{\pm,\text{reg}} = J_\pm \mp \left[\hat{\phi}_p - \frac{1}{\sqrt{2\upi}}\ln{\left(\frac{x^2+y^2}{1+y^2}\right)}\right]
\end{equation}
so that
\begin{equation}
\tilde{J}_{\pm,\text{reg}} = \tilde{J}_\pm \mp \left[\tilde{\phi}_p + \sqrt{2 \upi}\left(\frac{\ee^{-|k x|} - \ee^{-|k|}}{|k|}\right)\right].
\end{equation}
These variables are then indeed well-behaved (with no singularities) everywhere in Fourier and real space. In particular, for $k = 0$, 
\begin{equation}
\tilde{J}_{\pm,\text{reg}}(x,0) = \mp \left[\tilde{W}(x,0) + \sqrt{2 \upi}\left(1 - |x|\right)\right],
\end{equation}
where $\tilde{W}(x,0)$ may be found by evaluating (\ref{Wk0}). With these regularised variables, we are able to quickly achieve the earlier quoted absolute uncertainty in the flow solution of $1 \times 10^{-5}$.

\bsp	
\label{lastpage}
\end{document}